\documentclass[aps,prd,10pt,superscriptaddress,showpacs,floatfix,nofootinbib,eqsecnum,longbibliography,preprintnumbers]{revtex4-2}
\usepackage{color,graphicx,subfigure}
\usepackage{dcolumn,caption}
\usepackage{bm}
\usepackage{epsfig}
\usepackage{cancel}
\usepackage{slashed}
\usepackage[normalem]{ulem}
\usepackage{natbib}
\usepackage{amsmath}
\usepackage{amssymb}
\usepackage[bookmarksopen]{hyperref}
\usepackage{tabularx}

\def  \bea  {\begin{eqnarray}}
\def  \eea  {\end{eqnarray}}
\def  \nn   {\nonumber}

\def  \vecnabla   {\vec{\nabla}}
\def \vecB {\vec{B}}

\def \bq {{\bf q}}

\newcommand{\stkout}[1]{\ifmmode\text{\sout{\ensuremath{#1}}}\else\sout{#1}\fi}

\begin{document}

\title[]{Quantized electrical conductivity in binary neutron star mergers}

 \author{Sreemoyee Sarkar}%
 \email{sreemoyee.sinp@gmail.com}
\affiliation{Department of Physics, St.Xavier's College-Autonomous, Mumbai - 400001, India.
}%
 \author{Souvik Priyam Adhya}%
 \email{souvik.adhya@ifj.edu.pl}
\affiliation{%
Institute of Nuclear Physics, Polish Academy of Sciences, Radzikowskiego 152, 31-342 Krakow, Poland
}%
\date{\today}
\preprint{IFJPAN-IV-2021-12}

\begin{abstract}
 We examine nature of longitudinal electrical conductivity in magnetized electron-ion plasma in the context of binary neutron star mergers. In presence of strong magnetic field, high density and temperature, quantum oscillatory behavior for electrons emerge due to breakdown of the classical description. For pronounced thermodynamic effects, we consider zeroth Landau level population of electrons for electrical conductivity. We solve Boltzmann equation in presence of magnetic field to obtain the dissipative component of the conductivity. The conductivity is formulated considering dynamically scattering centers in the medium with magnetically modified screening. Numerical estimations show that the effect of magnetically modified screening mass on electrical conductivity is less. On the other hand, we observe that frequency dependent screening reduces electrical conductivity leading to a reduction in the Ohmic decay time scale to become of the order of the characteristic timescale of the merger process  in the low density regime. This indicates the relevance of dissipative process for the merger simulation in the above mentioned  domain.
 
\end{abstract}


\maketitle
\section{Introduction}
The recent detection of gravitational wave signal GW170817 originating from binary neutron star (BNS) merger by the LIGO and Virgo detectors have opened up a new era in multi-messenger astronomy \cite{LIGOScientific:2017vwq, LIGOScientific:2017zic, LIGOScientific:2017ync}. Additionally, short gamma-ray bursts (SGRBs) were also detected by the Fermi satellite GRB170817A indicating the presence of huge magnetic field in the merging event \cite{Kawamura:2016nmk, Paschalidis:2016agf, Ruiz:2016rai}. These mergers are unique astrophysical objects of significant sources of gravitational radiation, electromagnetic as well as neutrino emission \cite{Palenzuela:2013hu}. They offer a novel avenue to study highly non- linear gravitational effects blended with complex micro-physical processes; serving as Einstein's richest natural laboratory \cite{Baiotti:2016qnr}.

In the event post merging, a remnant neutron star is created and if the remnant possess a mass beyond Tolman–Oppenheimer–Volkoff (TOV) limiting mass, the merged object collapses within a few milliseconds. The description of neutron star mergers requires the knowledge of General Relativistic Magneto Hydro-dynamics (GRMHD)\cite{Kiuchi:2015qua,Kiuchi:2017zzg,Anderson:2008zp,Liu:2008xy,Dionysopoulou:2012zv,Dionysopoulou:2015tda,Ruiz:2017due,Palenzuela:2008sf, Palenzuela:2013kra}.  Most of these general-relativistic simulations account for ideal Magneto Hydro-dynamics (MHD) limit with infinite electrical conductivity($\sigma$) to arrive at large Ohmic decay time scale. In a recent work \cite{Harutyunyan:2018mpe}, the authors have pointed out that Hall decay time scale could  be relevant in the survival time period of merged object. In view of these recent studies \cite{Harutyunyan:2018mpe,Alford:2017rxf} on the relevance of different transport coefficients in BNS mergers, we analyse
 the importance of dissipative mechanism in the MHD simulation of mergers by evaluating quantized electrical conductivity with many-body effects and quantify the Ohmic decay time scale.
 
We consider  fully ionized plasma of electrons and ions . 
Heat and charge in  this medium are transported by electrons. The dominant electron transport mechanism is scattering on ions in the liquid phase. In presence of extreme magnetic field $(B \sim 10^{16} G)$ and density $(\rho \sim 10^{13} gm/cm^3)$, the classical
  description of electrons breaks down. Therefore, one should incorporate Landau quantization of energy levels in the formalism.  This quantization occurs for a particular set of temperature, density and magnetic field in case of neutron star. Thus, the inclusion of Landau quantization eventually modifies the non-magnetic electrical conductivity to great extent\cite{Yakovlev1980, Hernquist1984, Potekhin:1996zh, Potekhin:1996hu}. In the present paper, we focus mainly on the strongly quantizing case, since in this domain, the transport coefficients receive major modification due to the magnetic field. 
      
 The calculation of electrical conductivity by solving Boltzmannn equation in ultra-compressed plasma have been studied by several authors over the last few decades \cite{1964Abrikosov, 1966ApJ...146..858H, Lampe:1968zz, 1976ApJ...206..218F, 1976ApJ...206..218F,Schmitt:2017efp}. This requires the information of scattering rate of plasma constituents. The calculations of scattering rate considering screened Coulomb potential have already been observed in different Refs.\cite{1984MNRAS.209..511N, Yakovlev1980, Hernquist1984,Potekhin:1996zh, Potekhin:1996hu}. In all these calculations it has been assumed that ions are static scatterers. This formulation can not be easily transported to the relativistic domain of large densities where  dynamical effects could be important for reliable description of transport coefficients. Medium modified Hard-Thermal-Loop (HTL) and Hard-Dense-Loop (HDL) propagators include dynamical effects in the high temperature and high density plasmas  respectively \cite{Braaten:1989mz,Braaten:1990az,Altherr:1992mf,LeBellac:1996kr,Manuel:2000mk}. While Debye screening in plasma is related to the longitudinal photon exchange, the exchange of magnetic/ transverse photons contribute to dynamical screening of the plasma particles. 
 It is observed in different studies \cite{LeBellac:1996kr, Manuel:2000mk, Heiselberg:1993cr,Sarkar:2010bv,Sarkar:2011bc,Sarkar:2012ww,Adhya:2012sq,Adhya:2013ima} that for ultra-degenerate case, both in Quantum Chromodynamics (QCD) and Quantum Electrodynamics (QED) plasmas, the transverse
interactions not only become important but they dominate over their longitudinal interaction.   
In a recent calculation \cite{Harutyunyan:2016rxm}, the authors have included many-body effects through the HTL modified propagator in the calculation of  non- quantized electrical conductivity in a warm neutron star crust. Motivated by all these calculations of inclusion of dynamical screening in different transport coefficients, we include the medium modified propagator in quantized electrical conductivity in the context of BNS merger in the present paper. Here we perform the calculations of $\sigma$ in an extreme scenario of  temperature  $\sim 12 MeV$, density $\sim 10^{13}$ $gm/cm^{3}$ and magnetic field $\sim B_{16} G$. Finally, using the strongly quantized electrical conductivity, we estimate the Ohmic decay time scale and compare it with the survival time period of BNS mergers.

The paper is organised as follows. In section II, we present the physical conditions for relativistic and strongly quantized electrons in the BNS merger scenario. Next, in section III we derive the longitudinal electrical conductivity in a dynamically screened QED plasma. We study the Ohmic decay time scale of the magnetic field using the electrical conductivity as the input in section IV. We present numerical results for the electrical conductivity for typical ranges of temperature, magnetic field and density for the ultra-dense plasma along with estimation of Ohmic time scale in section V. Finally, we summarize and discuss the impact of dynamical screening on electrical conductivity and decay times in section VI. In addition, we include the important steps for the derivation of the conductivity in Appendix I.

\section{Physical Conditions}
Physical properties of  the BNS merger, which 
forms an unstable configuration are different
from isolated neutron stars.
 We consider simplest possible constituents of post-merger object of electron-ion plasma with  fully ionized ions and  free mobile electrons in the low density (up to $10^{12} gm/cm^3$), high magnetic field regime (up to $10^{17} G$ with $T\sim 15 MeV$).  Electron density $n_e$ is related to ion density $n_i$ {\em via} $n_e=Zn_i$ where $Z$ is the atomic number of the element. We consider the magnetic field ($B$) is present along the $z$ direction. Scattering of electrons with ions only contribute in electrical conductivity. 
 In the absence of magnetic field the electron density can be written as,
\begin{eqnarray}\label{elecdens1}
n_e&=&\frac{2}{(2\pi)^3}\int_0^{\infty}f(\epsilon) d^3p
\end{eqnarray}
 %
where, $f(\epsilon)=(exp(\frac{\epsilon-\mu}{kT}+1))^{-1}$. $\mu$ is the chemical potential written as $\mu^2=p_f^2c^2+m^2c^4$. 
In presence of a constant magnetic field, the electronic energy states are obtained as solutions of Dirac equations (\cite{Akhiezer, Lifshitz}). The positive energy states are denoted by quantum numbers $\epsilon, p_z,p_x,n,s$. $\epsilon$ is the electron energy, $p_z$ is the electron momentum along the field, $s =\pm 1$ is the helicity, and $n = 0,1,2$ enumerates the Landau levels. The energy of the relativistic electrons is given by,
\begin{eqnarray} \label{Disp}
 \varepsilon = \sqrt{p_z^2c^2 + m^2c^4 + 2n\hbar \omega_B mc^2},
 \end{eqnarray}
 instead of $ \epsilon=\sqrt{ p_f^2c^2+m^2c^4}$. Here, $\omega_B=ebB_{cr}/(mc^2)$ is the electron cyclotron frequency with $b=B/B_{cr}$ and $B_{cr}=4.413\times 10^{13}G$. 
 The ground Landau level is non-degenerate with respect to spin while the higher levels are doubly degenerate. 
 The number density of free electrons in presence of magnetic field is written as,
 \begin{eqnarray}\label{elecdens2}
 n_e=\frac{m\omega_B}{(2\pi)^2}\int_{-\infty}^{\infty}dp_z \sum_{n,s}f(\epsilon_n)
 \end{eqnarray}
 
 The magnetic field strongly quantizes the motion of electrons and different transport coefficients receive significant contribution when the electrons are confined to the zeroth Landau level. We do not consider the situation when ions receive quantum modifications due to the magnetic field. 
 Parameters which determine zeroth level population are as follows \cite{Chamel:2008ca},
 \begin{eqnarray}\label{TcerhoBpara}
&&T_{\mathrm{c}e}={\hbar \omega_{\mathrm{c}e}\over k_{\mathrm{B}}}\approx
1.343\times 10^8\;B_{10^{12}}~\mathrm{K}\nn\\
&&\rho_B=7.045\times 10^3{A\over
Z}\;({B_{10^{12}}})^{3/2} \mathrm{\ g\ cm}^{-3}
\label{temp_dens}
\end{eqnarray}
$B$ is strongly quantizing if $\rho<\rho_B$ and $T\ll T_{ce}$.

It is convenient to introduce the relativistic parameters $x_r=p_F/m_{ec} \sim 1.008(\frac{\rho_6 Z}{A})^{1/3}$ ($\rho_6=\rho/10^6$), $T_r=m_ec^2/k_B\sim 5.930\times 10^{9}$ K. The electron gas is relativistic for $x_r\gg 1$ or $T\gg T_r$. Hence, if density $\rho >10^{6}$ gm $cm^{-3}$  and temperature $T_r > 5.930\times 10^{9}$ K, electrons are relativistic corresponding to magnetic field $B=10^{14}$ G.
\footnote{We have used $c =k_B = \hslash = 1$, where $k_B$ is the Boltzmann constant and the electric charge $e$ is related to the fine structure constant by $\alpha = e^2/(4 \pi) = 1/137 $.}
\section{Formalism}
In the current paper, we consider fully ionized plasma of two components: electrons and positive ions of charge 
$Ze$. In case of compact objects, the huge magnetic field quantizes the motion of electrons present in the QED plasma.
 In this section we derive the expression for electrical conductivity ($\sigma$) in presence of strong magnetic field  in  electron-ion plasma from transport theory. 
 In presence of magnetic field $\sigma$ is anisotropic and the conductivity tensor is written as,
  \begin{equation}\label{sigmamatrix}
 \sigma = \Bigg(\begin{matrix} 
 \sigma_{\perp}&\sigma_{H}&0 \\ \sigma_{H}&\sigma_{\perp}&0\\0&0&\sigma_{\parallel}
 \end{matrix}
 \Bigg)
\end{equation}
where $\sigma_{H}$ is the Hall coefficient.
In the above expression, $\sigma_{\parallel}$ and $\sigma_{\perp}$ are the parallel and perpendicular components of $\sigma$ respectively in presence of external magnetic field along $z$ direction. In this section we present the calculation of quantized $\sigma_{\parallel}$ in electron ion plasma. For the rest of the paper, we re-define $\sigma_{\parallel}$ as $\sigma$.

$\sigma$ is related to the electric current density ($j$) and satisfies the constitutive relation  $j = \sigma E$ where $E$ is the electric field. We obtain $j$ from kinetic theory and is related to displacement of the electronic distributions from their equilibrium configuration due to the presence of electric field in the plasma. Hence,
 \begin{eqnarray}\label{ji}
 j&=&
 2e\int\frac{d^3 p}{(2\pi)^3} v\Phi
 \frac{\partial f_0}{\partial \epsilon}
\label{elec_current}
\end{eqnarray} 
in the above equation $e$ is the charge of an electron, $v=p/\epsilon$, $f_0$ is the equilibrium distribution function, $p$  and $\epsilon$ are the  the energy and the momentum of the particle.  In addition, $\Phi$ contains the information about the off-equilibrium distribution function which arises due to the presence of electromagnetic field in plasma. The $\Phi$ is obtained by solving Boltzmann equation in presence of magnetic field. In presence of small $E$, the distribution function evolves according to the magnetically modified Boltzmann equation given as \cite{Hernquist1984},
 \begin{eqnarray} \label{boltzmaneq}
\frac{\partial f_{np_zs}}{\partial t} + v_z\frac{\partial f_{np_zs}}{\partial z} - \dot{\textbf{p}}.\frac{\partial f_{np_zs}}{\partial p_z} =\mathcal{C}[f].
\label{be_b}
\end{eqnarray}
In the above equation, $ f_{np_zs}$ describes the population of electrons defined by the quantum state $n, s, p_z$. 
$v_z$ is the $z$ component of the velocity of the particle.
The third term in the LHS of eq.(\ref{boltzmaneq}) arises from the Lorentz force term $\dot{\textbf{p}}=e(\textbf{E}+\frac{1}{c}{\bf v_{p} \times B })$. In absence of external magnetic field, the Lorentz force term vanishes. In the Lorentz force, the magnetic field contribution vanishes ($ \frac{e}{c}(v_{p} \times B ).v_p \frac{\partial f_{np_zs}}{\partial \epsilon_{n,p_z,s}} =0$) and we obtain,
\begin{eqnarray}\label{lorforce}
\dot{\textbf{p}}.\frac{\partial f_{np_zs}}{\partial p_z}&\simeq&e\textbf{E}\frac{\partial f_{np_zs}}{\partial p_z } 
\end{eqnarray}
The RHS of Eq.(\ref{be_b}) contains the information of scattering rate of electrons with the ions present in the medium,
\begin{eqnarray}\label{Cf}
\mathcal{C}[f] = \frac{\partial f_{n,p_z,s}}{\partial t}\Bigg|_{coll}= \sum_f I_{f i}\big(f_{n'p_z' s'\rightarrow n p_z s} \big).
\end{eqnarray}
where, sum is over final state quantum numbers $n',p_z',s'$. $ I_{fi}$  is the electron-ion scattering rate from initial state ($i$) to the final state ($f$) in presence of $B$. $f_{n'p_z' s'}$ is the distribution function of the scattered state.
The distribution function of electrons has two parts in presence of electromagnetic field, 
\begin{eqnarray}\label{distfunc1}
f_{n, p_z, s} &=&f^{0}_{n,p_z,s}(\epsilon) + \delta f_{n,p_z,s}\\
f^{0}(\epsilon) &=& \frac{1}{\exp([\frac{\epsilon_{n,p_z, s}-\mu}{T}]+1)}
\end{eqnarray}
 where, $f^{0}(\epsilon)$ and $\delta f$ are the equilibrium and off- equilibrium distribution functions respectively.

We now proceed to calculate the collision integral considering strongly quantizing magnetic field. To calculate the interaction rate, we consider an electron with momentum $p = (E_{p},p_z, p_{\perp})$ and mass $m$ exchanges a virtual photon of momentum $q =$ ($q_0$, $q_z$, $q_{\perp})$ with an in-medium ion of momentum $k = (E_{k},k_z, k_{\perp})$ and mass $M$. The electron emerges with momentum $p' = (E_{p'},p'_z, p'_{\perp})$ and ion with momentum $k' = (E_{k'},k'_z, k'_{\perp})$. 
In order to obtain finite interaction rate, we use the HDL re-summed photon propagator with transverse and longitudinal contributions.

We start with the following expression of the interaction rate \cite{Hernquist1984},
\begin{eqnarray}\label{Ifi1}
&& I_{fi}=\frac{1}{2E}\int\frac{d^3p'}{(2\pi)^32E'}\int\frac{d^3k}{(2\pi)^32k}\int\frac{d^3k'}{(2\pi)^32k'}\nn\\
&& [f_{n, p_z, s} g_{k} (1-f_{n',p'_z, p'_{\perp}})-f_{n',p'_z, p'_{\perp}}g_{ k'}(1-f_{n,p_z, p_{\perp}})](2\pi)^4\delta^4(p+k-p'-k')|M_{fi}|^2
\eea
where $M_{fi}$ is the scattering matrix and given as
\cite{Harutyunyan:2016rxm},
\bea\label{eq:amplitude}
{\cal M}_{i\to f}=-\frac{J_0J'_0}{q^2-\Pi_L}+
\frac{\bm J_t\bm J'_t}{q^2-\Pi_T}=-{\cal M}_L+{\cal M}_T,
\label{mat_amp}
\eea
where, 
\bea\label{currents}
J^{\mu}&=&-e^*\bar{u}(p')\gamma^\mu u(p),\\
J'^{\mu}&=&Ze^*v_k^{\mu}=Ze^*(1, \vec{k}/M), 
\eea
are the components of currents. $e^{\star}=\sqrt{4\pi } e$, $Z$ is the atomic number of the nucleus and $v_k$ is the velocity of ion  with momentum $k$. The $\Pi_T$ and $\Pi_L$ HDL photon self-energies are transverse and longitudinal respectively. The form of the electronic spinors are given in the Appendix.

 To proceed further, we describe the screening mechanism of electron-ion plasma. In earlier calculations  (\cite{1964Abrikosov, 1966ApJ...146..858H, Lampe:1968zz, 1976ApJ...206..218F,1976ApJ...206..218F})  the authors have implemented static longitudinal component of photon propagator to  screen the Coulomb potential,
\begin{equation}
D_{\vec q} = \frac{1}{\vec q^{2}+m_{d}^{2}},
\end{equation}.
Following linear response theory, 
for time-dependent electric fields, there exists an additional screening mechanism, along with screened  Coulomb potential, related to the
energy transfer to the constituents of plasma known
as Landau damping. This arises  because of non-zero frequency of the plasma. We implement the effects of non-zero frequency in both the electric and magnetic components of the photon propagator  computed within the HTL/HDL  formalism. In principle in presence of magnetic field the photon propagator should be anisotropic. In the present study  we  consider the isotropic photon propagator and obtain,
\bea\label{eq:Dmunu}
D^{\mu \nu } (\omega, \vec{\mathbf{q}})=
\frac{P^{\mu \nu }(q)}{q^2{-}\Pi_T (q)} + 
 \frac{Q^{\mu \nu }(q)}{q^2{-}\Pi_L (q)}
\eea
where $q = (q_0,\vec{q})$ is the four-momentum of the photon and $P_{\mu\nu}$
and $Q_{\mu\nu}$ are the transverse and longitudinal projectors,
respectively, 
\bea\label{PQmunu}
P^{i j} (q) &=& -\delta^{ij} + \frac{q^i  q^j}{q^2},
\\
\qquad\qquad
Q^{00}(q) &=& -\frac{q^2}{\vec q^2} = 1 -\frac{q_0^2}{\vec q^2} = 1-y^2.
\eea
The transverse ($\Pi_T$) and longitudinal ($\Pi_L$) HDL photon
self-energies and are given by 
\bea
\label{PiT}
\Pi_T (q) &=& 3m_D^2 \left[ \frac{y^2}{2} + \frac{y (1{-}y^2)}{2} 
\ln\left(\frac{y{+}1}{y{-}1}\right)\right]\nn\\
\Pi_L (q) &=& 3m_D^2\left[ 1-y^2-\frac{y(1-y^2)}{2}\ln\left(\frac{y{+}1}{y{-}1}\right)\right],
\eea
where, $m_D^2= e^2dn_e/d\mu$. 

In presence of strong magnetic field electron density present in the plasma changes, leading to a modification in the screening. At low temperature and strong magnetic field  presence of sharp Fermi surface modifies the nature of screening. In the non-relativistic regime where $m\gg \mu$, for large B, the Debye mass is given by 
\begin{equation}
m_D^2 = 
(\frac{e}{\pi})^2(\frac{eB}{2})(\frac{m}{p_f^z})  \label{eq:mDnrel}\,,
\end{equation}	
and in the relativistic domain $m\ll \mu$, the Debye mass is given by 
\begin{equation}
m_D^2 = 
(\frac{e}{\pi})^2(\frac{eB}{2}) \label{eq:mDnrel}
\end{equation}	
 
In order to proceed further, we compute the phase space factor in the interaction rate given in eq.(\ref{Ifi1}). We neglect the  terms which are quadratic in distribution function as well as we do not consider the change of momentum of ions in the phase space factor. Hence, the phase term can be written as, 
\bea\label{phspacefac}
 [f_{n, p_z, s} g_{k} (1-f_{n',p'_z, p'_{\perp}})-f_{n',p'_z, p'_{\perp}}g_{k'}(1-f_{n,p_z, p_{\perp}})]=g_k(f_{n,p_z, p_{\perp}}-f_{n',p'_z, p'_{\perp}})
\eea
Thus the final expression for the interaction rate is obtained as (details in Appendix), 
\bea\label{Ifi2}
I_{fi}&=&\frac{n_i }{2}\sum_n\int  du(f_{n,p_z, p_{\perp}}-f_{n',p'_z, p'_{\perp}})[\frac{1 }{3(u+\frac{\xi}{3})(u+\xi)}-\frac{v_k^2}{6u(u+\frac{\xi}{3})}]{\cal F}
\eea
where, $n_i$ is the number density of ions. In order to find the transport coefficients, it is useful to define a
dimensionless scattering rate $a$ and a dimensionless perturbation to the distribution function ($f_1$) defined as,
\bea\label{Iandf1}\frac{I_{fi}}{n_iv_z\sigma_0}&=& a\nn\\
\frac{eE}{\sigma_0n_i}\frac{\partial f_0}{\partial \epsilon}\Phi&=&f_1
\eea
where,  $\sigma_0=\pi Z^2 e^4/ \omega_B^2$, $\omega_B=eB/m_e c$ . Using the above two equations, we obtain the dimensionless form of the linearized Boltzmann equation as shown below \cite{Yakovlev84},
\begin{equation}\label{eq:Beqdimless}
\sum_{n'~s'\gamma} a(n s \rightarrow n' s')\big(\Phi_{n' s'} - \gamma \Phi_{n s}  \big) = 1.
\end{equation}
$\gamma=\pm$ denotes the scattering channel for forward ($+$) and backward reactions ($-$). In the current paper, we present the results for the strongly quantizing scenario (i.e. zeroth Landau level)which provides the maximum effect with finite magnetic field in contrast to the  non-magnetic scenario. For zeroth Landau level, $n=n'=0$, spin degeneracy is absent and backward scattering ($\gamma=-1$) is the only allowed channel for scattering. Hence, after solving the dimensionless Boltzmann equation the off equilibrium distribution function is obtained as, 
  \bea\label{eq:phi}
 \Phi &=& \frac{E^{2} -1}{2Q_2} 
 \eea
 where, $E=\epsilon/mc^2$, $\Phi\equiv \Phi_{0,-1}$ and
\bea\label{eq:phi}
 Q_2 = \int_{0}^{\infty} e^{-u}[\frac{2}{3(u+\frac{\xi}{3})(u+\xi)}
 -\frac{v_k^2}{6u(u+\frac{\xi}{3})}] du
\eea
Finally, we obtain the expression for electrical conductivity for zeroth Landau level by inserting the value of $\Phi$
 in the equation below, 
\bea
j_z=\frac{em\omega_B}{4\pi^2}\int_{mc^2}^{\infty} f_1 d\epsilon
\eea
and then comparing with $j_z=\sigma E$.
Thus the final form of $\sigma$ becomes,
\begin{eqnarray}\label{eq:Q2}
\sigma&=&\frac{ m^4 b^2}{8\pi^3 Z^2e^2n_i}
\int_{mc^2}^{\infty}\frac{\partial f_0}{\partial \epsilon}\Phi d\epsilon
\label{sigma_final}
\end{eqnarray}
The above expression can be written in more compact form by introducing energy dependent electron relaxation time $\tau(\epsilon)$,
\bea\label{eq:sigma0LL}
\sigma&=&e^2\int_{mc^2}^{\infty}\frac{{\cal N(\epsilon)}\tau(\epsilon)}{\epsilon}(-\frac{\partial f}{\partial \epsilon})d\epsilon
\eea
where, $\tau(\epsilon)$ is                                      \bea\label{eq:sigmareltau}
\tau(\epsilon)&=&\frac{\epsilon l m \omega_B \Phi}{2\pi ^2 {\cal N(\epsilon)}}\nn\\
{\cal N}(\epsilon)&=&\frac{2m\omega_B}{4\pi^2}p_{n=0}(\epsilon)
\eea
$l$ is the electron scattering length $l=\frac{mc^2\omega_B}{2\pi n_i Z^2e^{4}}$.                                                                                                         

\section{Ohmic decay time scale}
The estimation of Ohmic dissipation time  scale
 is important to assess the
 limit of the ideal MHD approximation. Ideal MHD  is  defined
to be the limit in which the electrical resistivity $\eta = 1/\sigma$ vanishes. Using Maxwell's equation one can obtain the magnetic field decay time scale as described below. We start with Ampere's and Faraday's law,
\bea
{\vec {\nabla}} 
    &\times&\vec B=\frac{4\pi \vec j}{c}\nn\\
  {\vec {\nabla}} 
    &\times& \vec E=-\frac{\partial \vec B}{\partial t}.
    \label{maxwell_eq}
\eea
Now, $j=\sigma E$ and resistivity $\rho=c^2/4\pi \sigma^{-1}$. Using these two in the above Eq.(\ref{maxwell_eq}) Ohm's law can be written as follows by eliminating electric field.
\begin{eqnarray}
 && \frac{\partial \vec B}{\partial t}=
  -{\vec {\nabla}\times} ~\left(\vec{\hat{\varrho}}  \, \, {\vec{ \nabla}
    \times}{\vec B}\right)\,.
\end{eqnarray}
 The above equation is temed as the induction equation. Using vector identity ${\vec{\nabla} \times}\, ({\vec{\nabla} \times}\,\vecB) =\vecnabla\,
(\vecnabla\cdot\vecB)-\nabla^2 \vecB$
and $\vecnabla\cdot\vecB=0$, we obtain from Eq.~\eqref{maxwell_eq},
\begin{eqnarray}
  \label{eq:mag_field_is1}
  \frac{4\pi\sigma}{c^2}\frac{\partial \vecB}{\partial t}=\nabla^2\vecB
  .
\end{eqnarray}
A qualitative estimate of the magnetic field decay timescale ($\tau$) can be obtained from Eq.~\eqref{eq:mag_field_is1} if we
approximate $|\nabla^2 \vecB|\simeq B/\lambda_B^2$ and
$|\partial \vecB/\partial t|\simeq B/\tau$, where, $\lambda_B^2$ is the characteristic length scale of variation
of the magnetic field.  From these estimates,
we find that the magnetic field decay (or diffusion) timescale due to Ohmic dissipation is given by the well-known expression \cite{Harutyunyan:2017lrm, Rezzolla_book:2013},
\begin{eqnarray}
  \label{eq:decay_time}
\tau = \frac{4\pi\sigma \lambda^2_B}{c^2}\, .
\end{eqnarray}
In the next section, we  study the effects of the frequency dependent screening on $\tau$ entering through the  $\sigma$ as defined in the above equation.
\section{Results and discussions}
In this section we describe the behaviour of longitudinal quantized electrical conductivity with density, temperature, magnetic field and atomic number for the hot and dense electron- ion plasma. In the merger scenario, the electrons are considered relativistic for $T_{tr} > 5\times 10^9$ K and density $\rho\sim 10^6$ gm $cm^{-3}$. The momentum of an electron is related to the energy via the
relation $p_{nz}/(mc^2) =\sqrt{(\epsilon/mc^2)^2-2bn-1}$. From this expression one can obtain the
maximum Landau level that the electrons can populate and is given by the
integer part of $\nu=(E^2-1)/2b$ ($E=\epsilon/mc^2$). The energy of the electrons is constrained to $(E^2-2b) < 1$ to meet the condition of lowest Landau level. This is an important condition to be
used to obtain the desired results for the plots of $\sigma$ as we describe later
in this section. The parameters for density, temperature and magnetic field are appropriately chosen for relativistic quantized electrons to simultaneously meet the physical conditions applicable for the merging scenario.

\subsection{Variation with atomic number}
 In fig.\ref{fig:1}, we have plotted $\sigma$ with $\rho$ for two different equation of states (EOS): BPS (Baym, Pethick and Surtherland) \cite{Baym:1971pw} and magnetic BPS model \cite{Nandi:2010fp}. In ref.\cite{Baym:1971pw} the equation of state of zero-temperature matter in complete nuclear equilibrium is given for mass densities below $5\times 10^{14}$ $g cm^{-3}$.  In Ref.(\cite{Nandi:2010fp}), the BPS \cite{Baym:1971pw} has been extended to include the physical parameters for a low density plasma in presence of high magnetic field relevant for neutron star crust. We choose $Mo$ for calculations of $\sigma$ at higher density as shown in fig.\ref{fig:1}. In addition, we have also presented results for $Fe$ as a reference. Thus, we consider these two elements for the estimation of $\sigma$. 

\subsection{Effect of magnetically modified electronic screening}
In the figs.\ref{fig:2a} and \ref{fig:2b} we have shown the variation of $\sigma$ with density for $Mo$ and $Fe$ respectively. For the current plot we consider magnetically modified  static screened Coulomb potential. We consider two different sets of temperature. It can be seen that the effect of screening is negligible in electrical conductivity. Next we introduce the medium modifications through the HDL corrected photon propagator for the calculation of $\sigma$. 

 \subsection{Effect of HDL modified propagator}
We have plotted $\sigma$ with $\rho$ in Fig.\ref{fig:3a} and Fig.\ref{fig:3b} for different temperatures and at a fixed magnetic field with the HDL modifications through the interaction rate. We find that the inclusion of HDL propagator in the calculation reduces the value of $\sigma$ substantially. 

It is known that modifications to different equilibrium and non- equilibrium properties of plasma due to inclusion of HDL propagators emerge from frequency dependent transverse component of the propagator. However in case of $\sigma$, inclusion of the transverse photon does not provide a significant change as the transverse term contains a pre- factor which is inversely proportional to the mass of the ions. The reduction in $\sigma$ arises due to increase of the interaction rate caused by non- zero component of the plasma frequency.

  \subsection{Variation with density}
 Figs.\ref{fig:4} shows the variation of $\sigma$ with $\rho$ for different temperatures and at fixed magnetic fields. In order to consider electrons to be relativistic, the density and temperatures are chosen as $\rho \gg 10^{6} gm/cm^3$ and $T\gg T_r$ respectively. 
 For fixed $B$, $\mu$ increases with $\rho$ and electrons start to populate higher Landau levels. Since, we are interested in population of the zeroth Landau level, the density and temperature should satisfy $\rho<\rho_B$ and $T\ll T_{ce}$ as defined in Eq.(\ref{temp_dens}). 
 With all these conditions, $\sigma$ has been obtained by numerically integrating the expression Eq.(\ref{sigma_final}).
  The variation of $\sigma$ with $\rho$ in the fig. (\ref{fig:4}) shows that at temperature $\le 10^{10}$ K, a prominent hump is present. This nature occurs  because at this density the weak degeneracy condition ($|\epsilon-\mu|\sim T$) of electron distribution function is fulfilled. The nature of the curve resembles differentiated Fermi function at $T \ll \mu$. As the temperature increases, the hump gets flattened since electrons start becoming non-degenerate.
\subsection{Variation with temperature and magnetic field}
Fig.(\ref{fig:5}) shows variation of $\sigma$ with $T$ for different densities. The plot shows hump at particular temperature where the density is such that the condition $(e-\mu)\sim T$ is satisfied. 
The $\sigma$ in fig.(\ref{fig:5}) can be fitted as,
\bea\label{eq:abcparam}
\sigma&=&(a+b\times T^c)^{-1}\nn\\
a&=&1.45\times 10^{-25}\nn\\
b&=&1.05\times 10^{-46}\nn\\
c&=&1.919
\eea
At low temperature the effect of $T^c$ is very small, hence, $\sigma$ is constant. On the other hand at high temperature $\sigma\propto  T^{-c}$ and decreases with temperature. Thus, at higher temperatures, the electrons become classical obeying the inverse dependence of temperature. 
Fig.(\ref{fig:6})shows the variation of $\sigma$ with magnetic field $B$ for different densities. The fitting parameters for the $\sigma$ with $B$ is obtained as,
\bea
&&\sigma=A_0+A_1B+A_2B^2\nn\\
&&A_0=5.15\times 10^{27}\nn\\
&&A_1=-2.71\times 10^{11}\nn\\
&&A_2=5.30\times 10^{-6}
\eea
The peak in the left panel occurs at $\epsilon - \mu(B) \sim T$. On increasing the magnetic field, $\sigma$ increases with $B$ and saturates. For higher temperature (right panel), $\sigma$ shows a gradual increment with $B$.
\begin{figure}
\centering
\includegraphics[width=80mm]{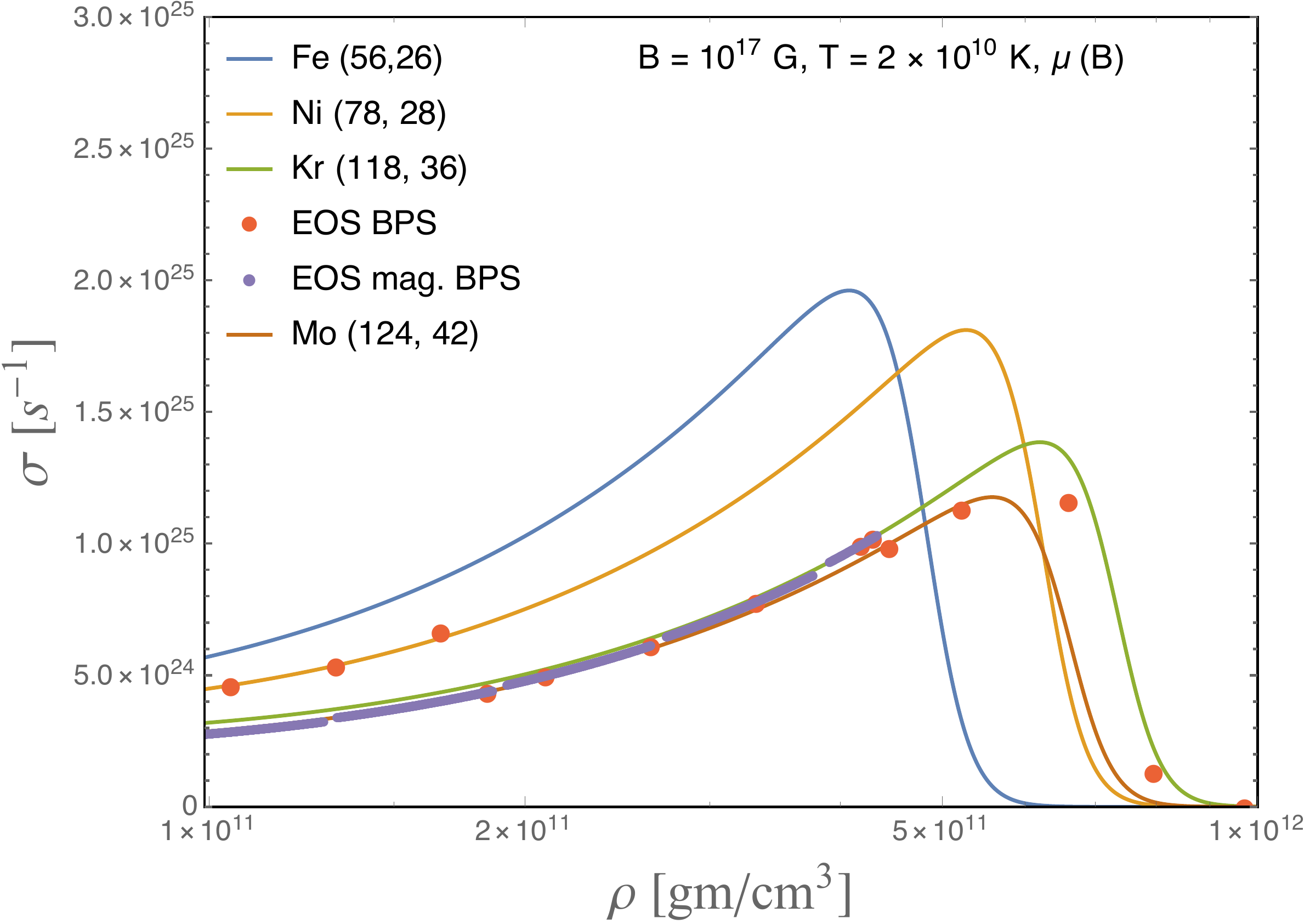}
\caption{The comparison of $\sigma$ with $\rho$ for different elements. The magnetic field is chosen as $10^{17} G$ and temperature as $2 \times 10^{10} K$. We have shown the comparison with the EOS {\cite{Baym:1971pw, Nandi:2010fp}}}
\label{fig:1}
\end{figure}
\begin{figure}
\centering     
\subfigure[$Mo$ $(124, 42)$]{\label{fig:2a}\includegraphics[width=80mm]{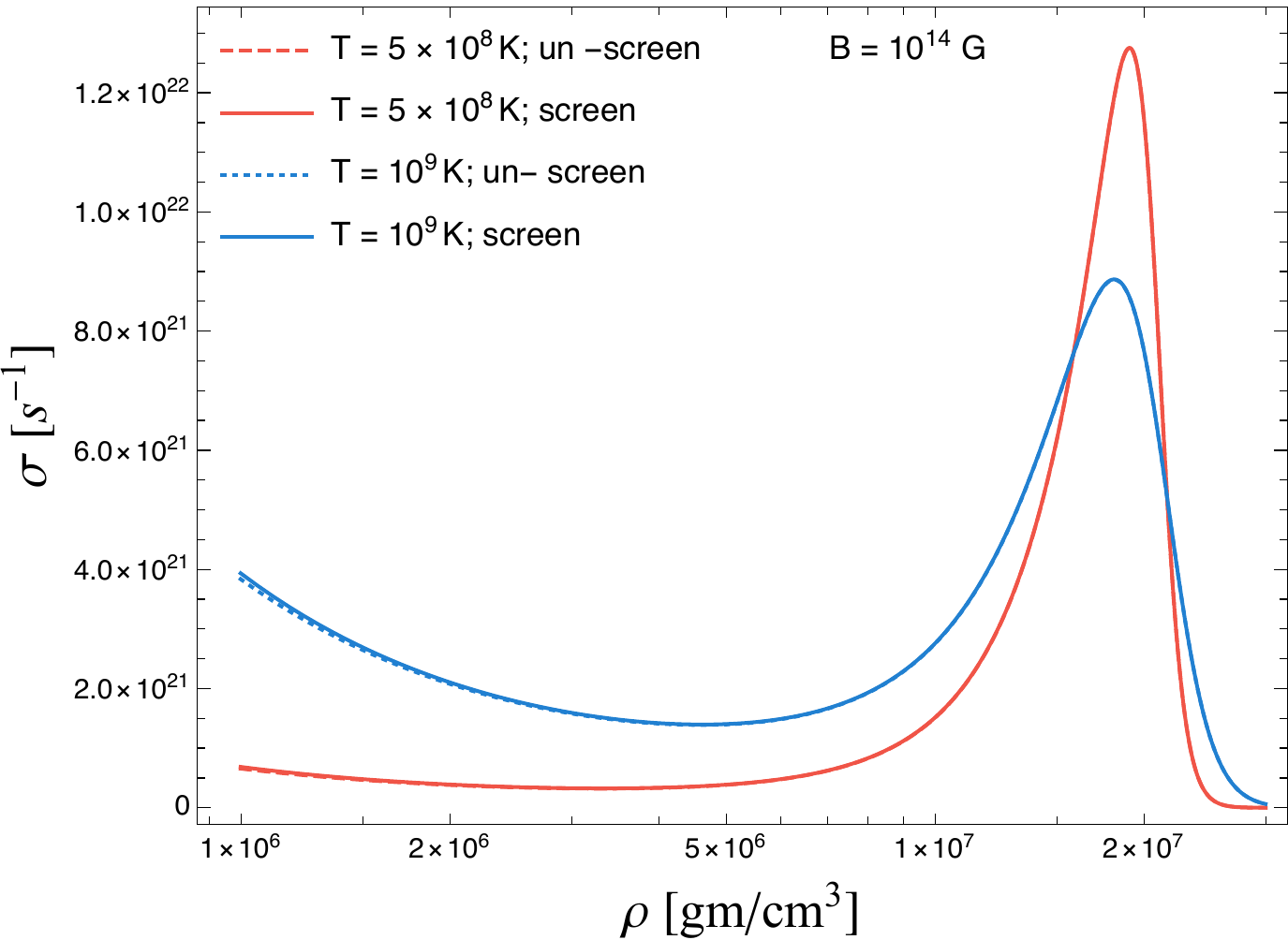}}
\subfigure[$Fe$ $(56, 26)$]{\label{fig:2b}\includegraphics[width=80mm]{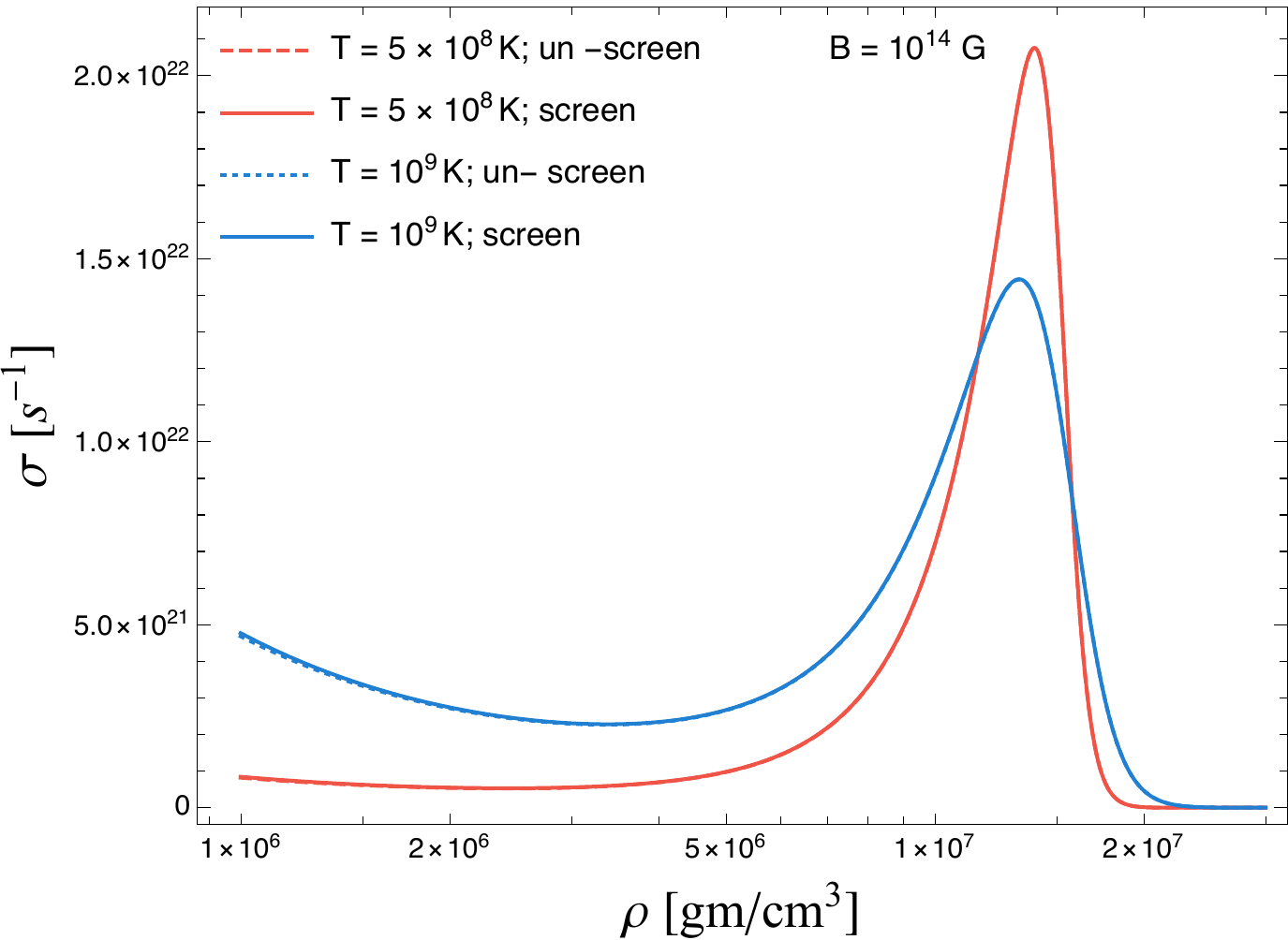}}
\caption{The comparison of $\sigma$ with $\rho$ for magnetically modified Debye screening and non- magnetic Debye screening cases for different values of temperature at a fixed field of $10^{14} G$. The choice of elements are $Mo$ $(124, 42)$ (left panel) and $Fe$ $(56, 26)$ (right panel) respectively.}
\label{fig:2}
\end{figure}
\begin{figure}
\centering     
\subfigure[$Mo$ $(124, 42)$]{\label{fig:3a}\includegraphics[width=80mm]{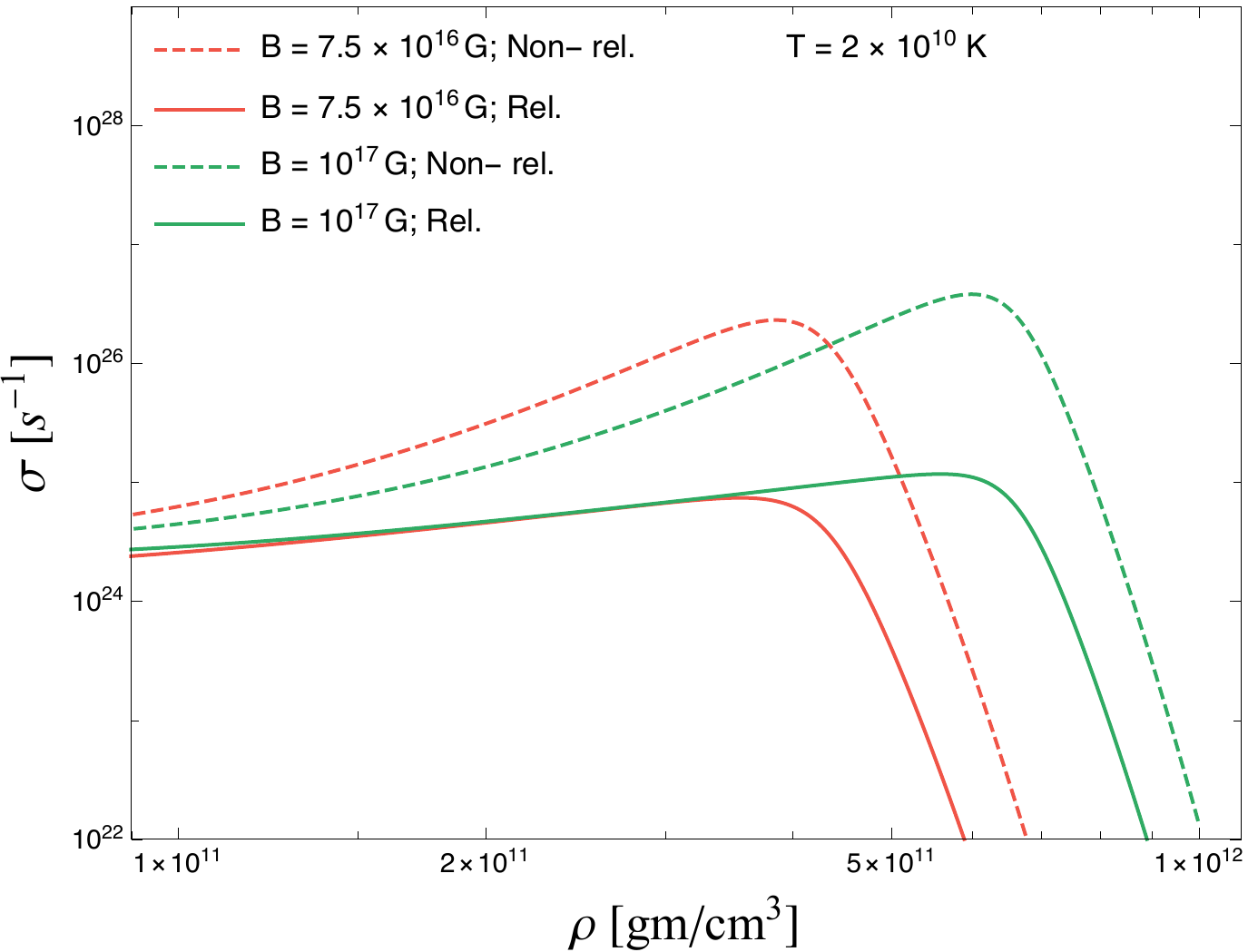}}
\subfigure[$Fe$ $(56, 26)$]{\label{fig:3b}\includegraphics[width=80mm]{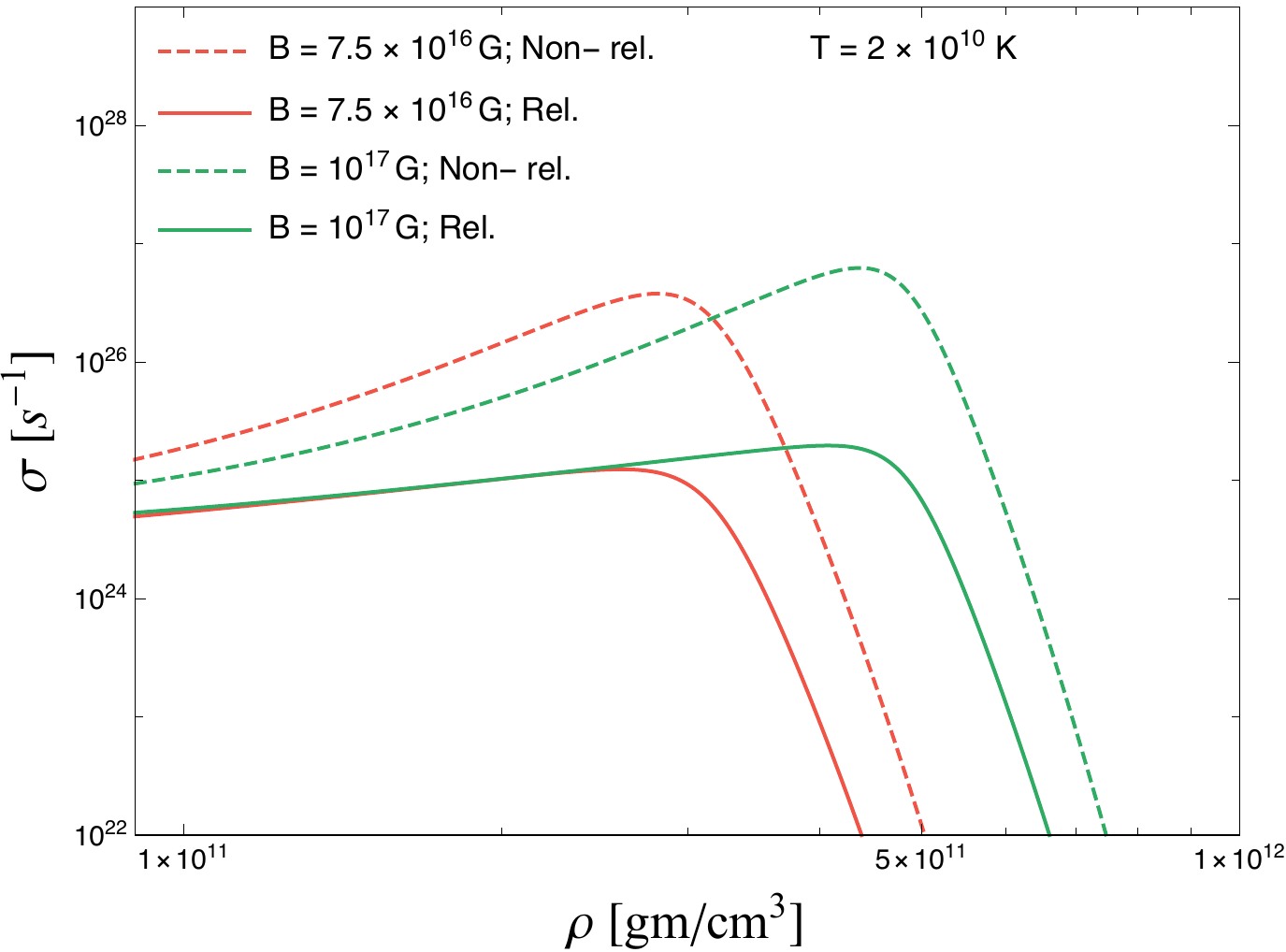}}
\caption{The comparison of $\sigma$ with $\rho$ considering relativistic and non- relativistic photon polarisation functions for different values of magnetic field at a fixed temperature of $2 \times 10^{10} K$. The choice of elements are $Mo$ $(124, 42)$ (left panel) and $Fe$ $(56, 26)$ (right panel) respectively.}
\label{fig:3}
\end{figure}
\begin{figure}
\centering     
\subfigure[$Mo$ $(124, 42)$]{\label{fig:4a}\includegraphics[width=80mm]{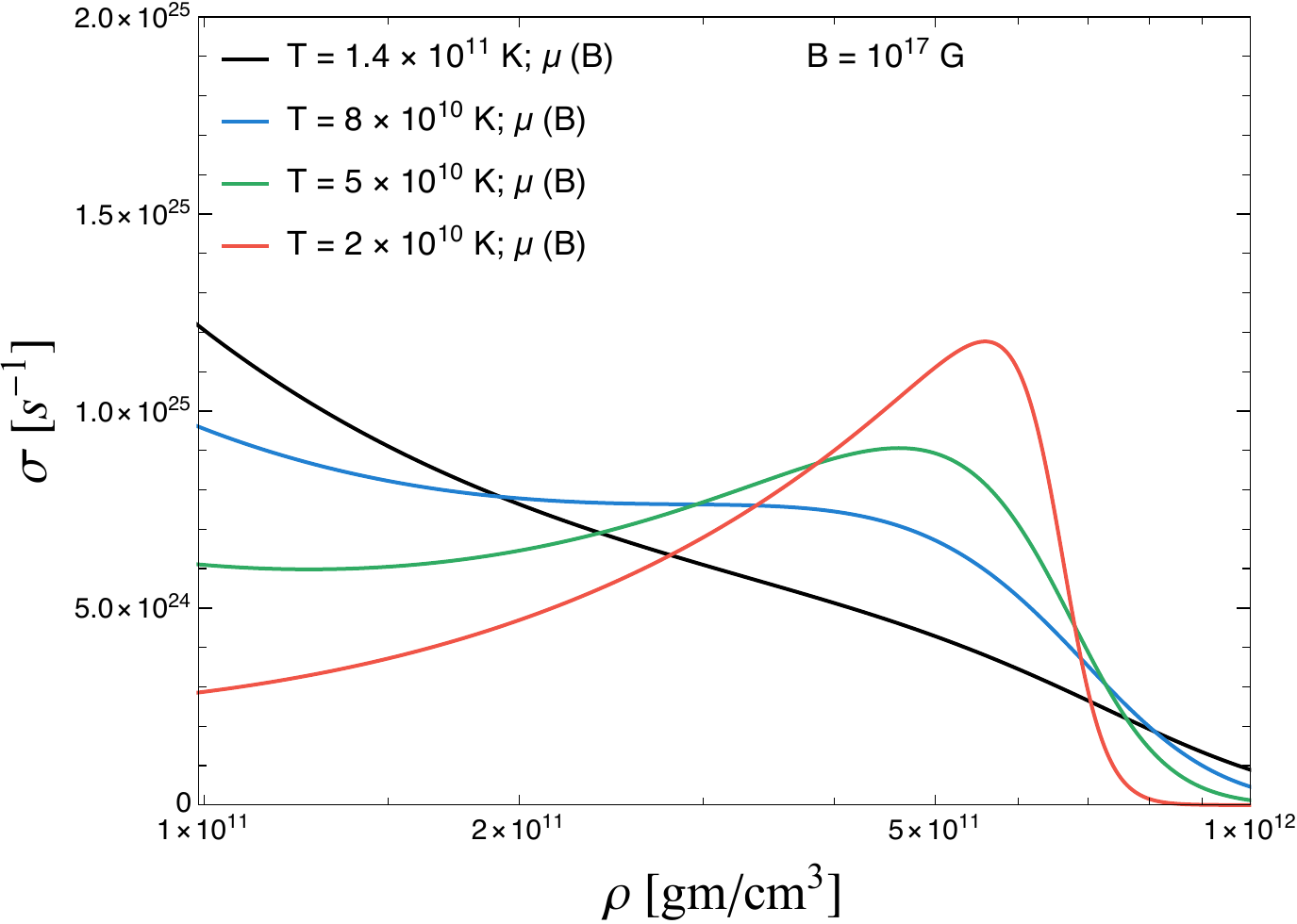}}
\subfigure[$Mo$ $(124, 42)$]{\label{fig:4b}\includegraphics[width=80mm]{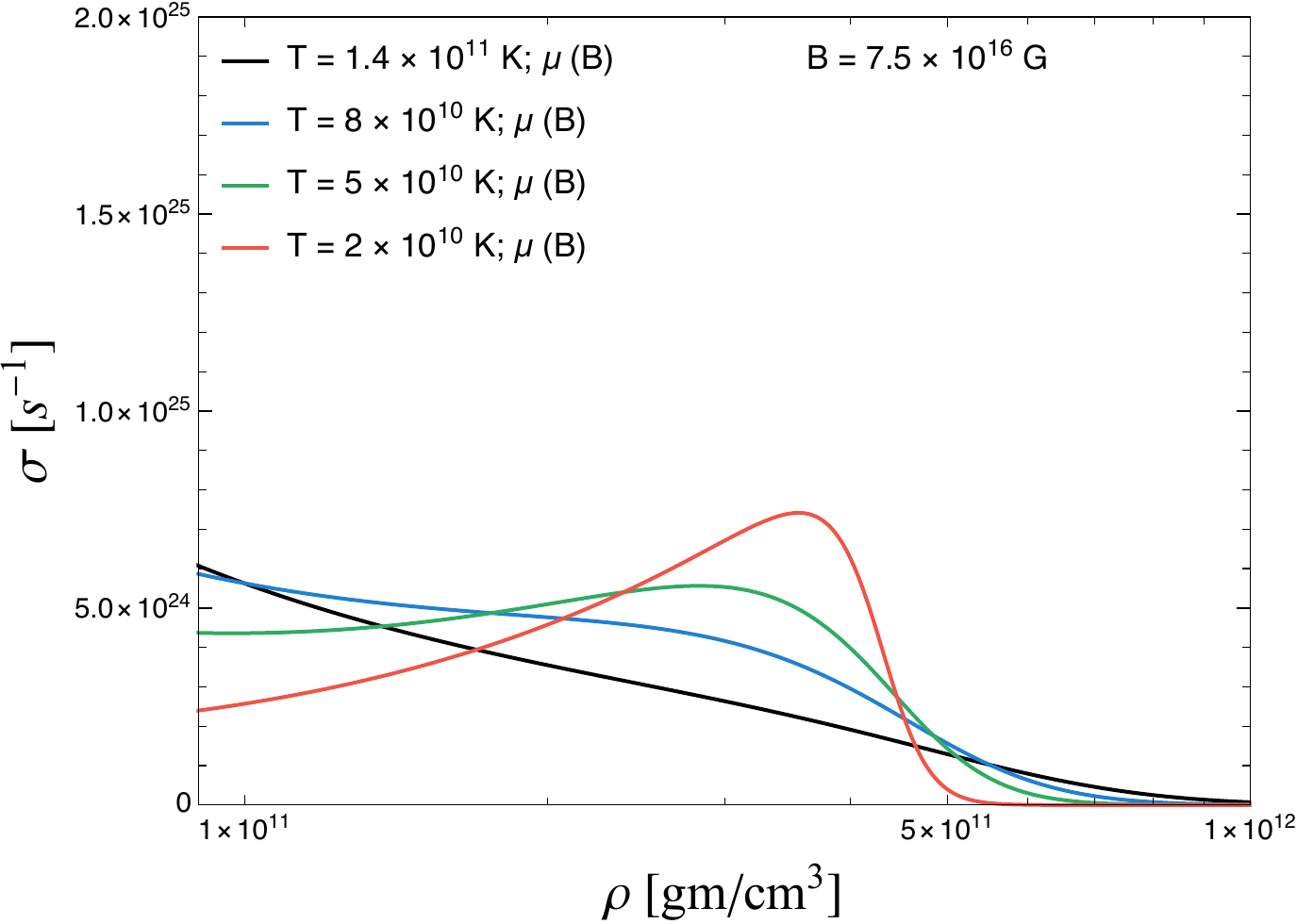}}
\subfigure[$Fe$ $(56, 26)$]{\label{fig:4c}\includegraphics[width=80mm]{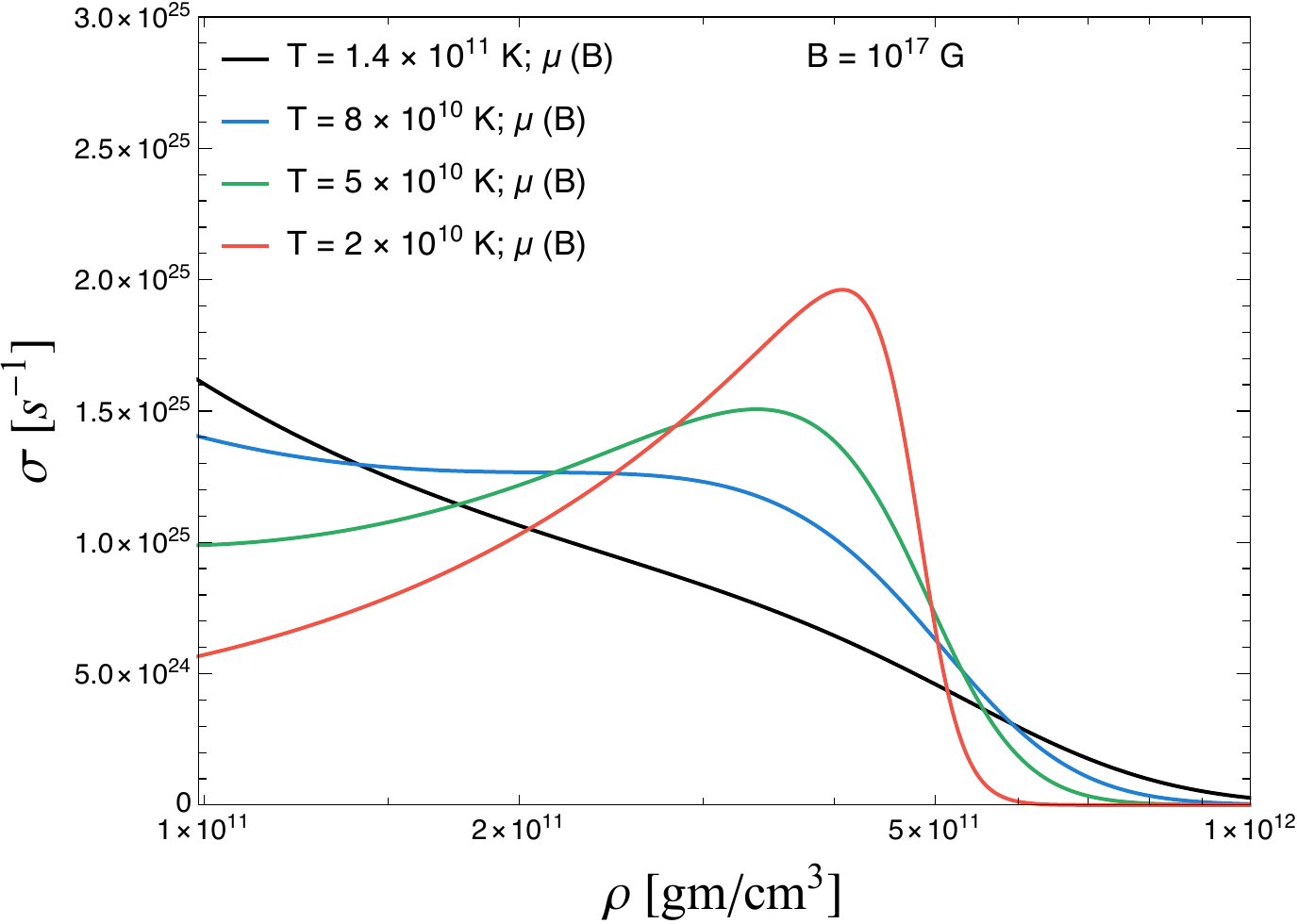}}
\subfigure[$Fe$ $(56, 26)$]{\label{fig:4d}\includegraphics[width=80mm]{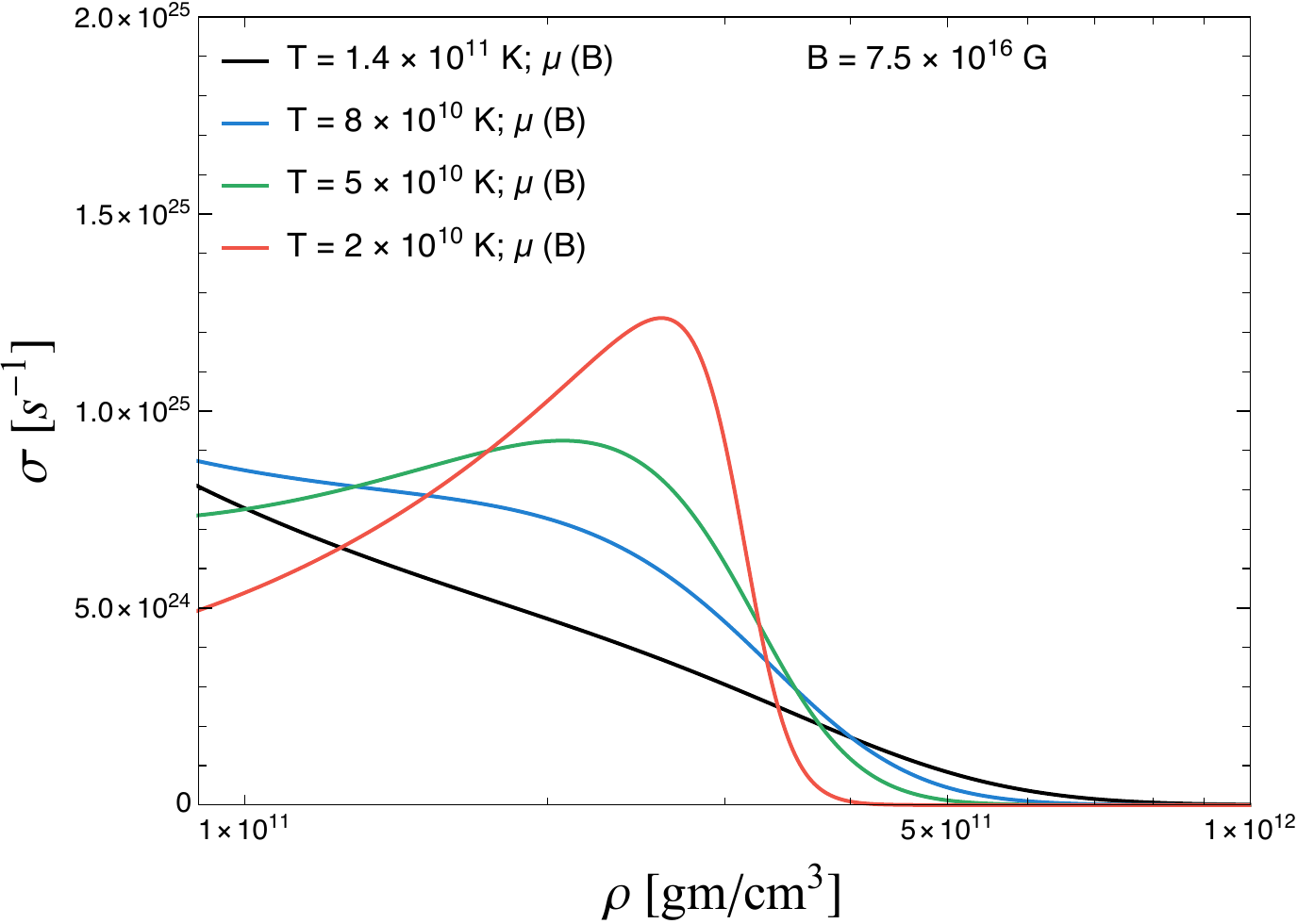}}
\caption{The comparison of $\sigma$ with $\rho$ for different values of temperature. The magnetic fields chosen are $10^{17} G$ (left panel) and $7.5 \times 10^{17} G$ (right panel) respectively. The choice of elements are $Mo$ $(124, 42)$ (upper panel) and $Fe$ $(56, 26)$ (lower panel) respectively.}
\label{fig:4}
\end{figure}
\begin{figure}
\centering     
\subfigure[$Mo$ $(124, 42)$]{\label{fig:5a}\includegraphics[width=80mm]{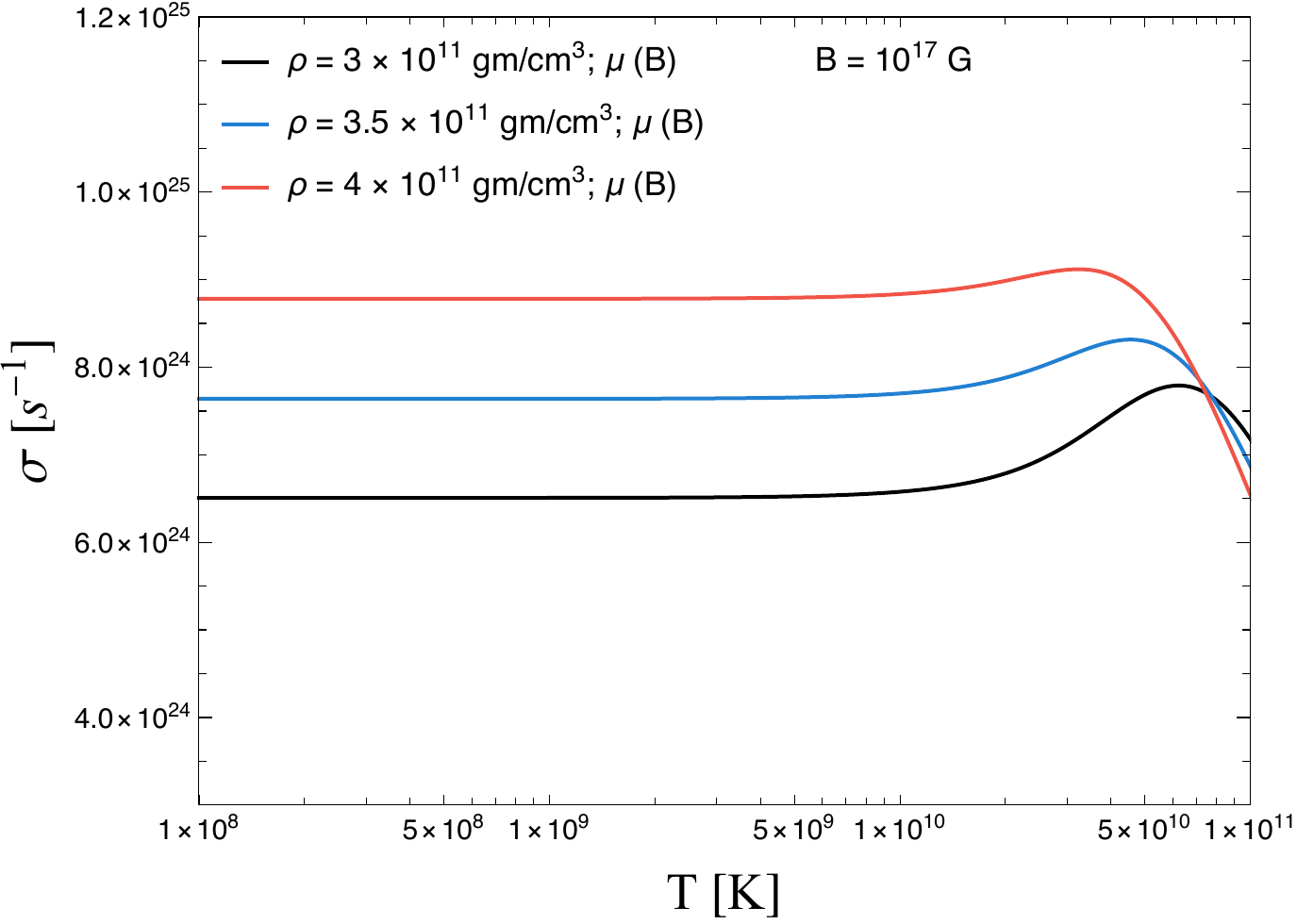}}
\subfigure[$Mo$ $(124, 42)$]{\label{fig:5b}\includegraphics[width=80mm]{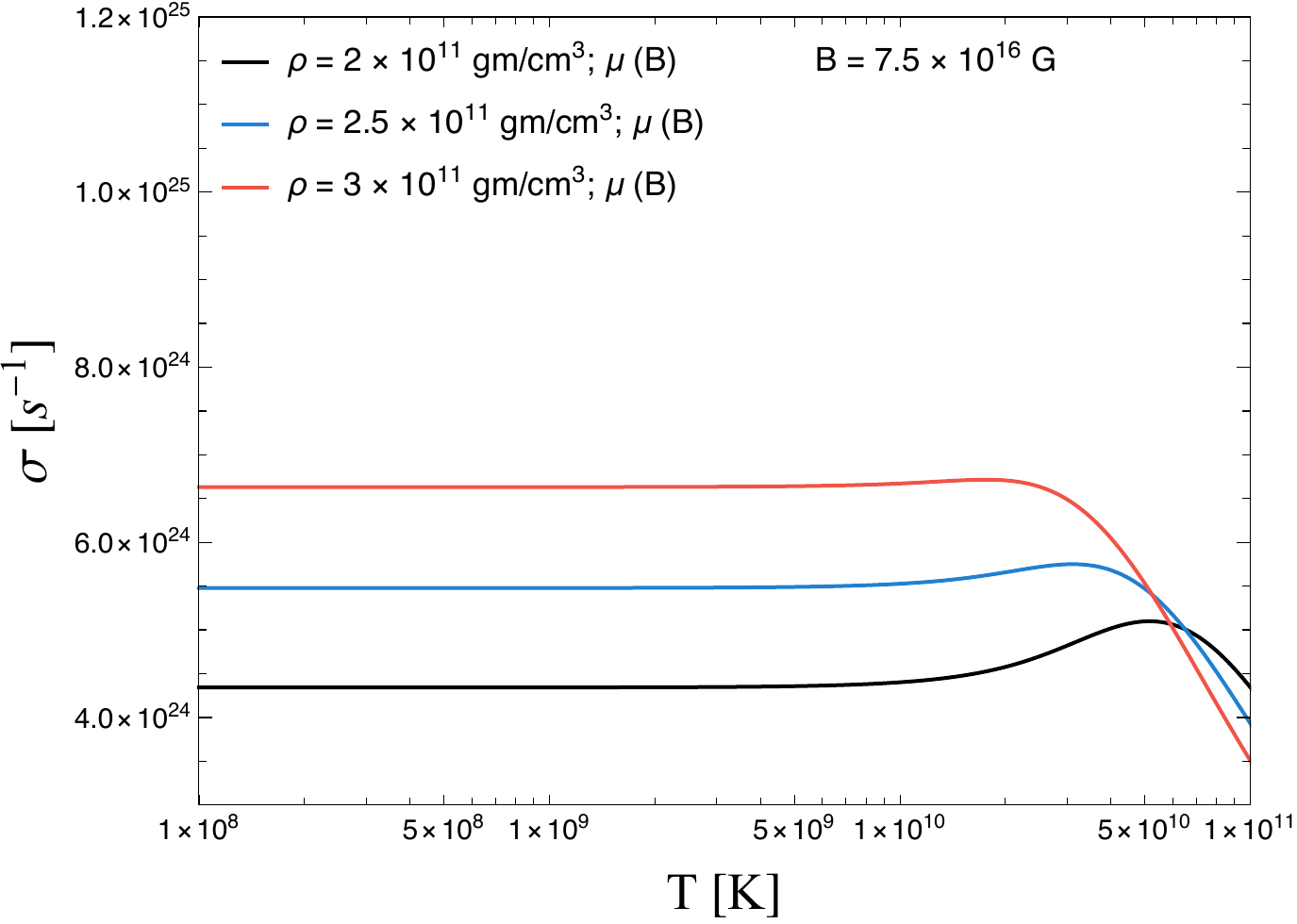}}
\subfigure[$Fe$ $(56, 26)$]{\label{fig:5c}\includegraphics[width=80mm]{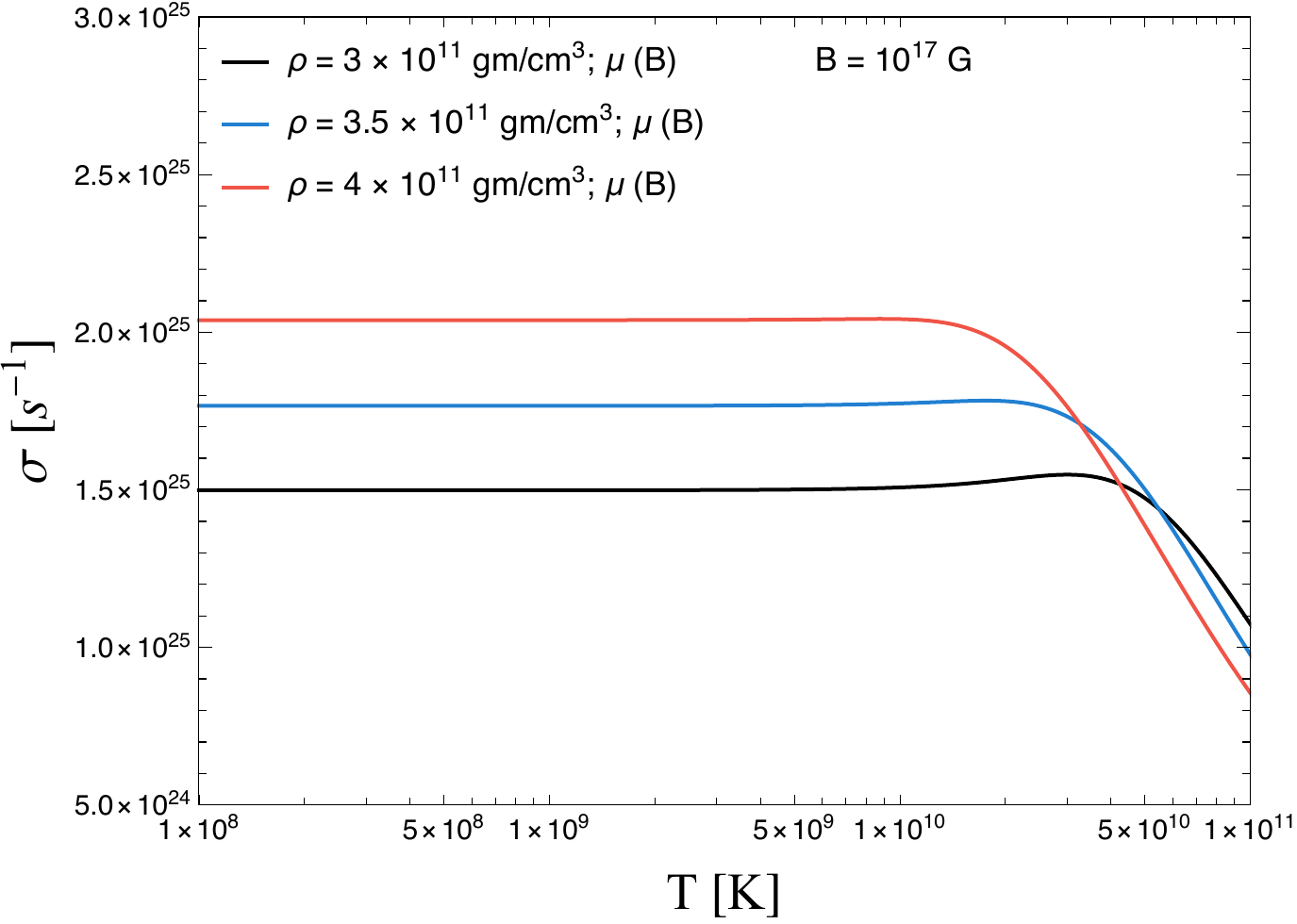}}
\subfigure[$Fe$ $(56, 26)$]{\label{fig:5d}\includegraphics[width=80mm]{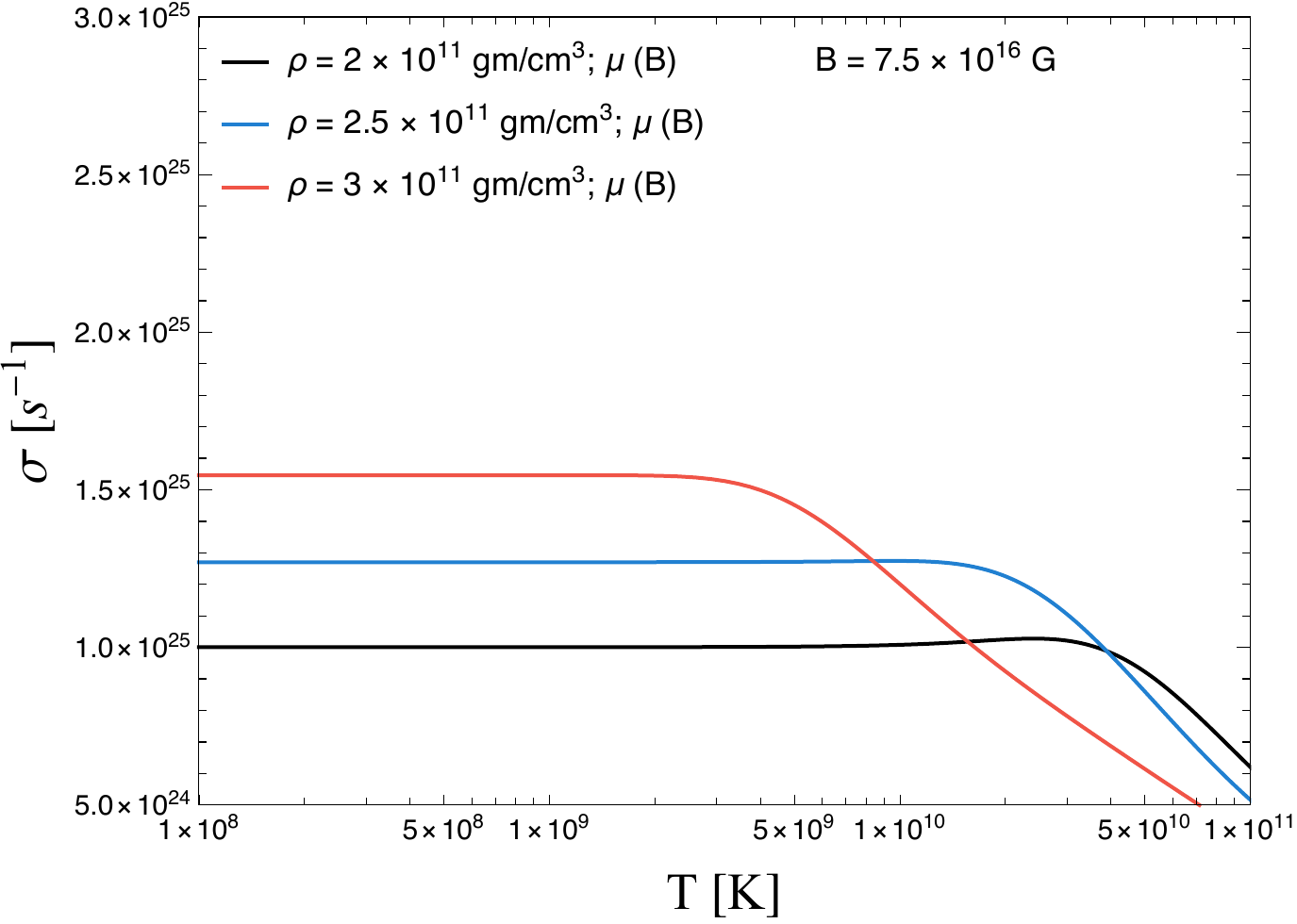}}
\caption{The comparison of $\sigma$ with $T$ for different values of magnetic field. The fields chosen are $10^{17} G$ (left panel) and $7.5\times10^{17} G$ (right panel) respectively. The choice of elements are $Mo$ $(124, 42)$ (upper panel) and $Fe$ $(56, 26)$ (lower panel) respectively.}
\label{fig:5}
\end{figure}
\begin{figure}
\centering     
\subfigure[$Mo$ $(124, 42)$]{\label{fig:6a}\includegraphics[width=80mm]{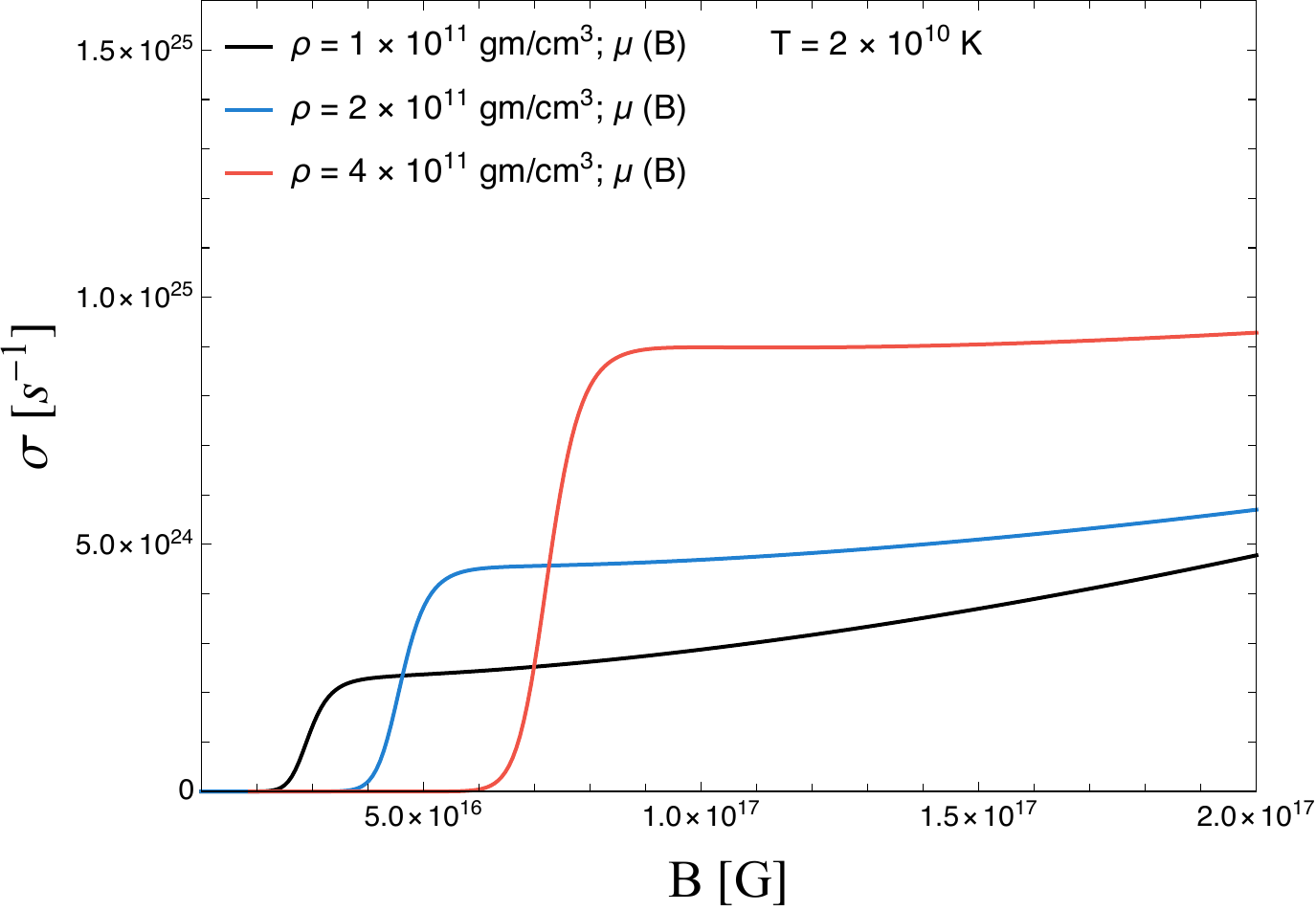}}
\subfigure[$Mo$ $(124, 42)$]{\label{fig:6b}\includegraphics[width=80mm]{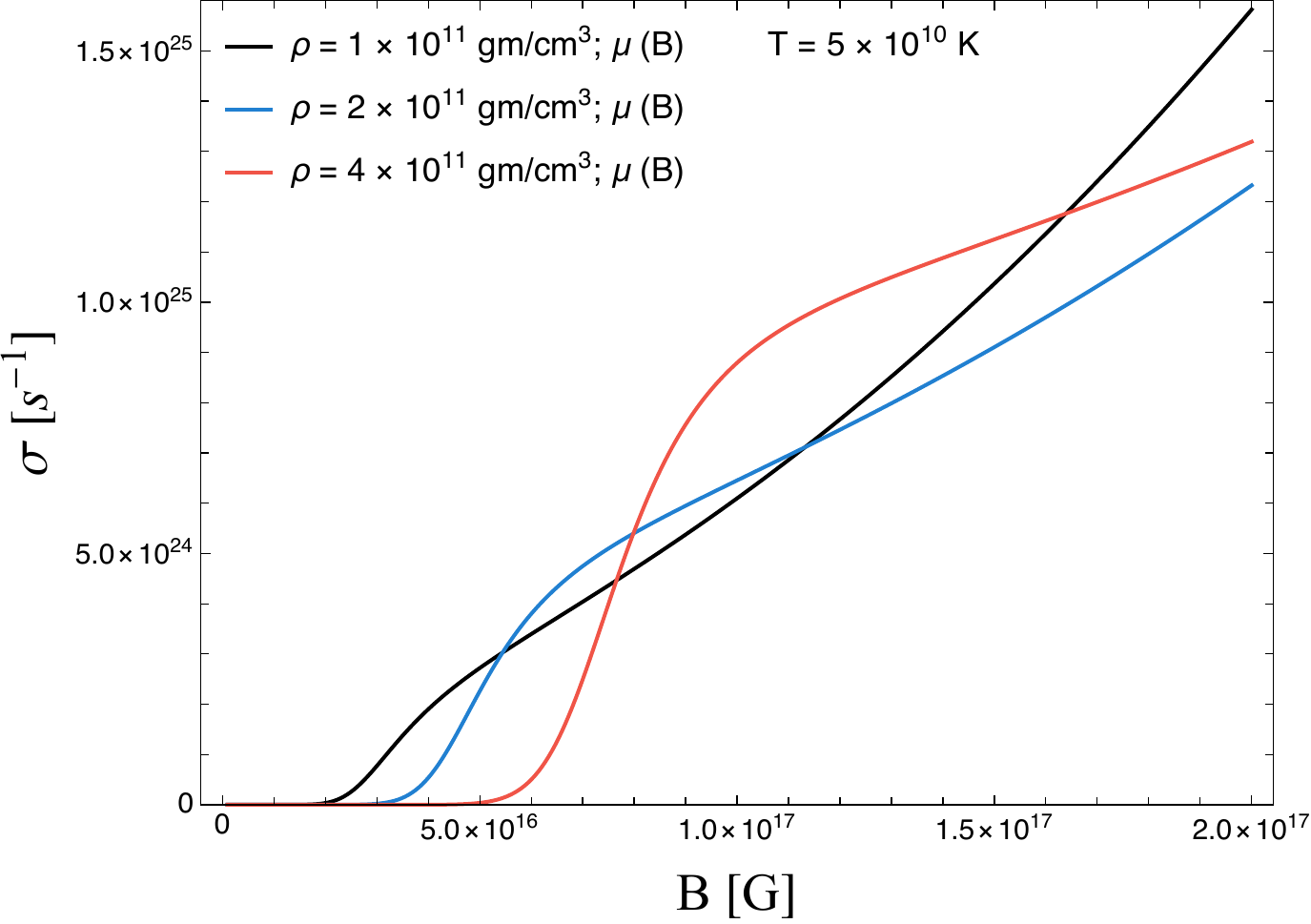}}
\subfigure[$Fe$ $(56, 26)$]{\label{fig:6c}\includegraphics[width=80mm]{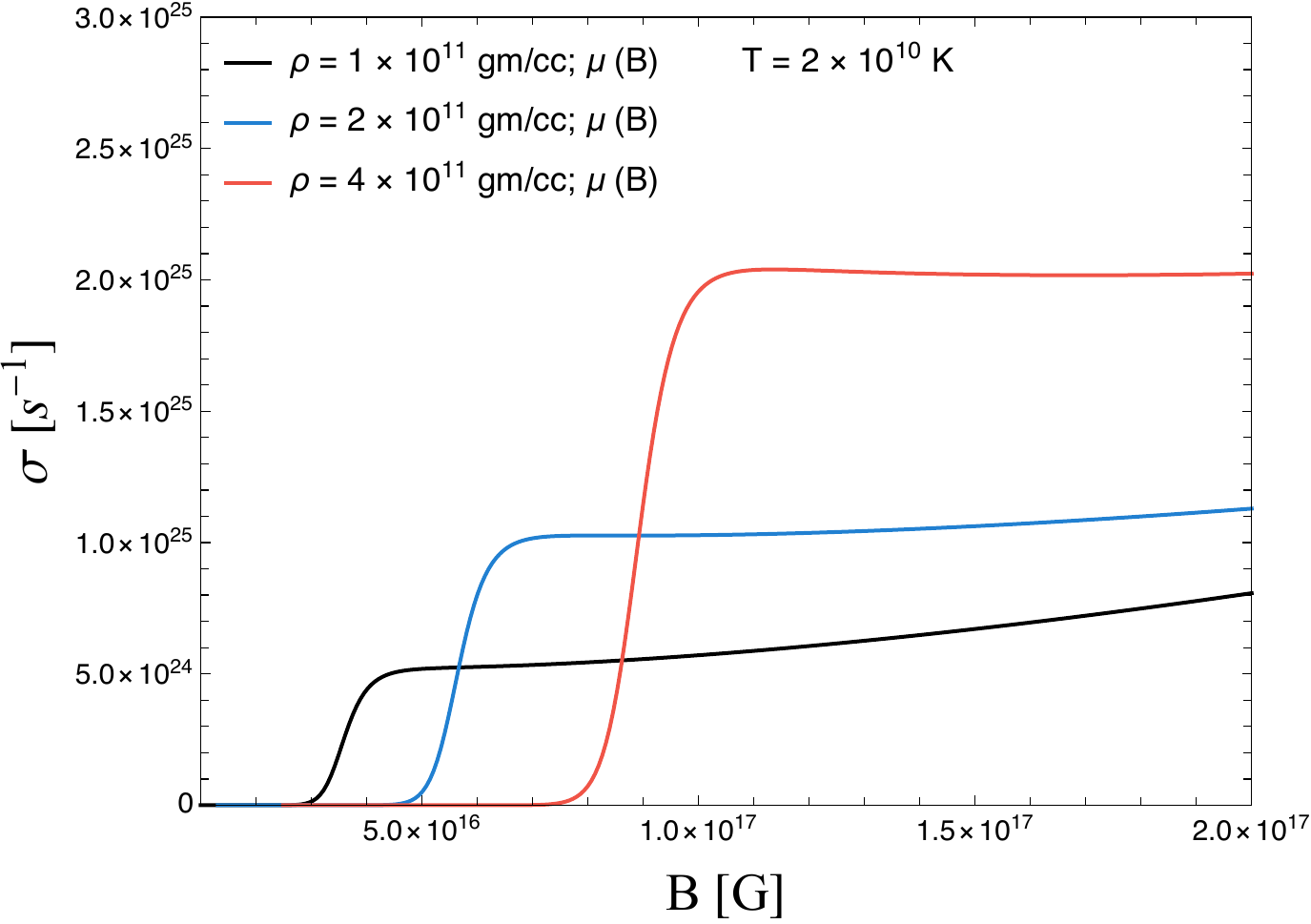}}
\subfigure[$Fe$ $(56, 26)$]{\label{fig:6d}\includegraphics[width=80mm]{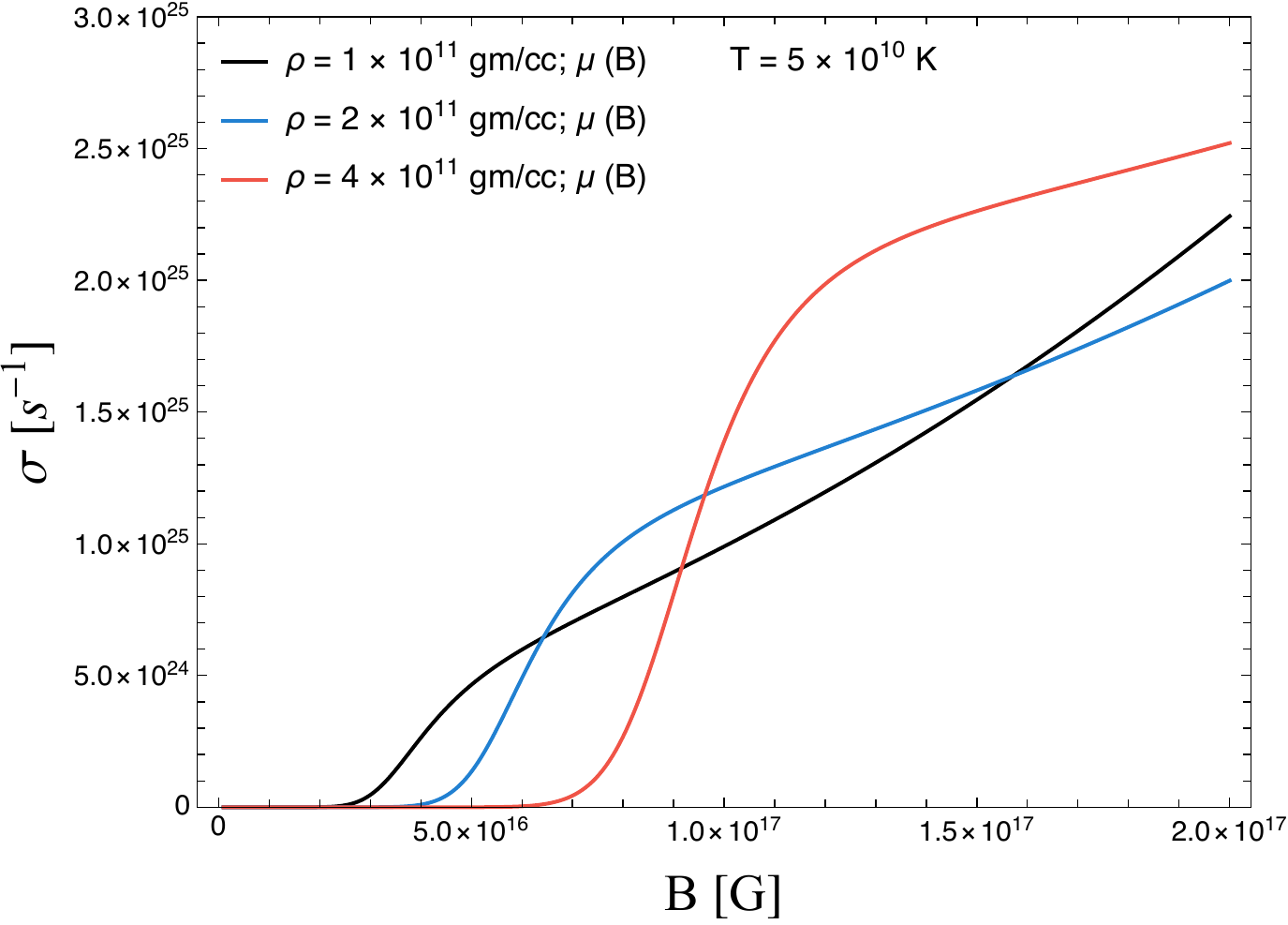}}
\caption{The comparison of $\sigma$ with $B$ for different values of density. The temperatures chosen are $2\times10^{10} K$ (left panel) and $5\times10^{10} K$ (right panel) respectively. The choice of elements are $Mo$ $(124, 42)$ (upper panel) and $Fe$ $(56, 26)$ (lower panel) respectively.}
\label{fig:6}
\end{figure}
\subsection{Estimation of Ohmic time scale}
The decay timescale of magnetic field due to the Ohmic dissipation is given by the expression  Eq.(\ref{eq:mag_field_is1}). For the estimation of the time scale, the information of average scale height of magnetic field is required. The average scale is defined as $\lambda_B\equiv B/\nabla B$ and a limiting value has been assigned to it following $\lambda_B \simeq 0.1\times10^{-5}\left(T/1 MeV\right)^{-1/2}m$ \cite{Harutyunyan:2018mpe},
In fig.\ref{fig3D:1}, we present the estimation of the Ohmic decay timescale for a typical range of density and temperature for two values of the magnetic field. We find the timescale $\sim 1-2$ $ms$ which is well within the range of survival time period of neutron star merger. However, the naive estimation presented here demands a more realistic and detailed derivation of different time scales to assess the validity of MHD simulation in neutron star mergers. 
\begin{figure}
\centering     
\subfigure[$Mo$ $(124, 42); B = 7.5 \times 10^{16} G$]{\label{fig3D:1a}\includegraphics[width=80mm]{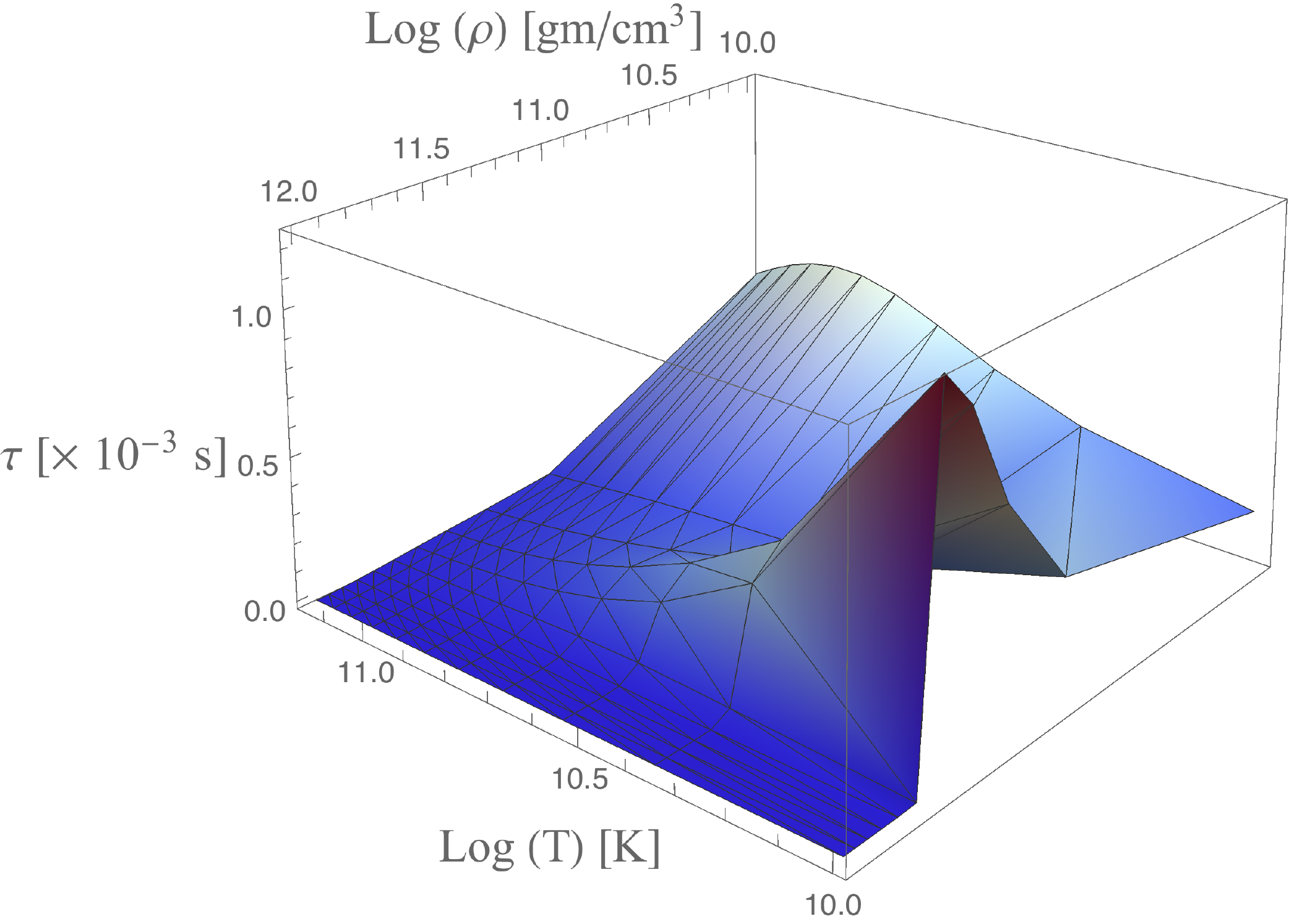}}
\subfigure[$Mo$ $(124, 42); B = 10^{17} G $]{\label{fig3D:1b}\includegraphics[width=80mm]{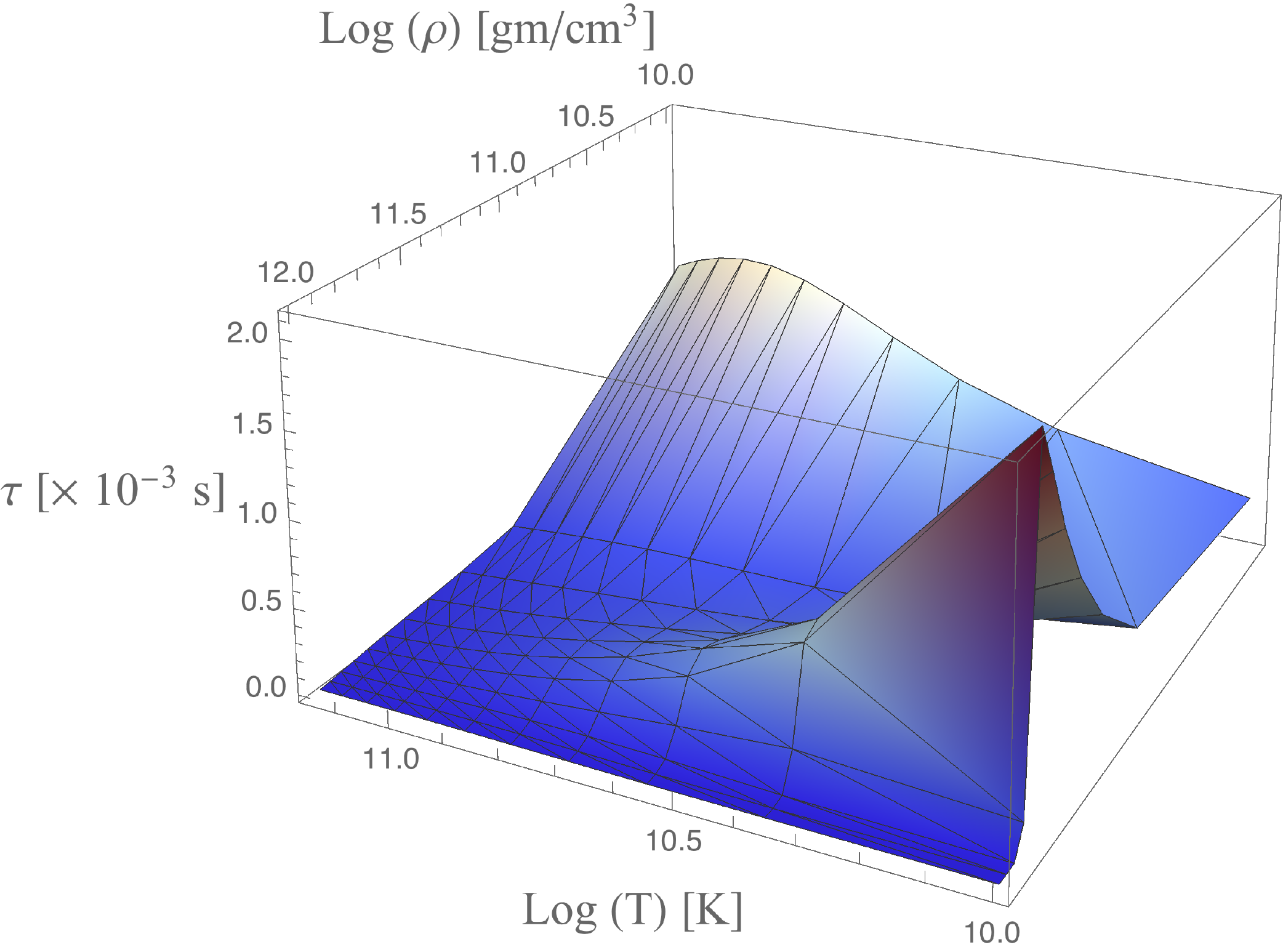}}
\caption{The estimation of $\tau$ with $\rho$ and $T$. The magnetic field is chosen as $10^{17} G$ and $7.5 \times 10^{16} G$ respectively for $Mo$ $(124, 42)$.}
\label{fig3D:1}
\end{figure}
\section{Summary and conclusions}
In this work, we have calculated the quantized longitudinal electrical conductivity in presence of magnetic field followed by numerical results which account for the inclusion of the dynamical screening  in the calculation. The calculation has been performed considering electron-ion plasma where the dominant mechanism is the scattering of electrons with ions through screened electromagnetic force. We have presented the plots of variation of electrical conductivity with density ( $\sim 10^{12}$ $gm/cm^3$), temperature ($\sim 10^{10}$ $K$) and magnetic field ($\sim 10^{17}$ $G$) considering two metals $Mo$ and $Fe$. The scales for the generation of the plots are chosen so that they obey primarily the conditions that the electrons become relativistic at density $\rho>10^{6}$ $gm/cm^3$ and temperature $T> 5.93\times 10^{9}$ K and electrons remain confined to zeroth Landau level by obeying the condition $\sqrt{E^2-1}/2b\ll 1$. Considering these two constraints, the domain of validity of our calculations lie in the high magnetic field and low density plasma of BNS mergers.

For the calculation of conductivity, we have assumed particles are slightly out of equilibrium which allows us to solve the Boltzmann equation numerically. We have considered electron-ion scattering amplitude via screened electromagnetic interaction using magnetically modified spinors. The off-equilibrium distribution function has been obtained by solving the Boltzmann kinetic equation in relaxation time approximation. However, we have not considered the finite size of the nuclei and ion structure function for the calculation of the relaxation rate.

The electromagnetic interaction between electrons and ions have been incorporated through HDL propagators in the calculation. The calculation should account for magnetically modified anisotropic  photon propagator; however for the present paper, we have considered only isotropic HDL propagator and aim to include magnetically modified propagator in an upcoming work \cite{aritra2021}. We have observed that the effect of magnetically modified Debye screening is negligible in electrical conductivity. 
 
We have found that the inclusion of the HDL propagator reduces the value of $\sigma$ in contrast to static screening. The transverse component of the propagator contributes less to the conductivity since it is inversely proportional to the mass of the ion.
The frequency dependent screening in the longitudinal component of the HDL propagator enhances the interaction rate thereby causing a decrease in $\sigma$.
Next, we have estimated the Ohmic decay time scale from the expression of quantized $\sigma$ including frequency dependent screening. The obtained time scale is of the same order as the survival time of the merged compact object. Hence, one can infer that many-body effects play an important role in determining the dissipative time scales relevant in the neutron star merger.
In the current paper, we have considered only longitudinal electrical conductivity (and ignored all the other tensorial components) in the equation for magnetic-field evolution in plasma.
Hence, the realistic estimate of $\tau$ can only be obtained if all the components of conductivity tensor are known in the background of relevant equation of state for neutron star merger. However, our calculations presented in this paper provide a significant step towards conjoining the many body effects in plasma with observational hydrodynamic simulations.

\subsection*{Acknowledgments}
SS would like to thank DST- INSPIRE faculty scheme Award no.: DST/INSPIRE/04/2015/002594 during the tenure of which this work was initiated and UM-DAE CEBS for hosting the INSPIRE project. Authors would also like to thank and acknowledge Rana Nandi for providing the EOS data as well as fruitful discussions regarding various aspects of this work. Authors acknowledges fruitful discussion with Tanay Mazumdar regarding  numerical analysis of the current project.
SPA would like to acknowledge IPNP, Charles University, Prague where the initial stages of the work was carried out.
\section*{Appendix I}
\label{App}
In this Appendix, we present the important steps for the evaluation of electron-ion scattering rate in presence of magnetic field. An electron of momentum $P(\epsilon_p, \vec {p})$ scatters with an ion of momentum $K(\epsilon_k, \vec {k})$ leading to the final momentum states $P'(\epsilon_{p'}, \vec {p'})$ and $K'(\epsilon_{k'}, \vec {k'})$. The scattering rate from initial state  to final state is given by,
\bea\label{eq:Ifi1}
I_{fi} &=& \frac{1}{4}\int\frac{d^3p'}{(2\pi)^32}\int\frac{d^3k}{(2\pi)^32}\frac{d^3k'}{(2\pi)^32}
 [n_f(p)g_f(k)(1-n_f(p'))-n_f(p')g_f(k')(1-n_f(p))]\nn\\&& (2\pi)\delta(\epsilon_p+\epsilon_k-\epsilon_{p'}-\epsilon_{k'})\delta^3(\vec {p}+\vec {k} -\vec {p'}-\vec {k'})|{\cal M}_{fi}|^2
 \eea
The above equation can be modified in presence of non-zero magnetic field as,
\bea\label{eq:Ifi3}
I_{fi}&=&\frac{eB}{(2\pi)^2}\sum_n\int dp'_zdy_B\frac{d^3k}{(2\pi)^3}\frac{ dq_y }{16}\int\frac{d\Omega_k}{4\pi}g_f(k)[n_f(p)-n_f(p')]\delta(\epsilon_p+\epsilon_k-\epsilon_p'-\epsilon_k')\nn\\ &&\sum_{q_x,q_z}\delta_{k_x'-k_x,q_x}\delta_{k_z'-k_z,q_z}|{\cal M}_{fi}|^2
\label{int_rate2}
\eea
where, $y_B=\frac{p_x}{m\omega_B}$ and we have inserted $\int d\Omega_k/4\pi=1$.
The argument of delta function can be written as follows,
\bea\label{eq:deltaeps}
\delta(\epsilon_p+\epsilon_k-\epsilon_{p'}-\epsilon_{k'})&=&\delta(\epsilon_{k}-\epsilon_{k'}-\hat p.\vec{q})
\eea
where, we have used $\epsilon_{|p'|}=\epsilon_{|p-q|}=\epsilon_p-\hat p.\vec{q}$. The angular integrals of $\vec{k}$ can be expressed as,
\bea\label{eq:angint1}
&&\int \frac{d\Omega_k}{4\pi}\delta(\hat p.{\bf q}-v_k.{\bf q})=\frac{1}{2q}\nn\\
&&\int \frac{d\Omega_k}{4\pi}\delta(\hat p.{\bf q}-v_k.{\bf q})(\hat p.\hat k-\hat p.\hat q \hat q.\hat k)^2=\frac{1}{4q}\left(1-\frac{(\hat p.{\bf q})^2}{q^2}\right)\left((v_k^2-\frac{(\hat p.{\bf q})^2}{q^2}\right)
\eea
 Eq.(\ref{int_rate2}) can be written as,
\bea\label{eq:Ifi4}
I_{fi}&=&\frac{n_i}{(2\pi)^2 }\sum_n\int dq_zdq_x\frac{ dq_y }{16\times 2q}g_f(k)[n_f(p)-n_f(p')] |{\cal M}_{fi}|^2
\eea
where, we have changed the variable $dy_B'=\hbar dq_x/m\omega_B$ following momentum conservation $p'_x-p_{x}=q_x$. In the above equation, $n_i$ is the number density of ions  which can  be expressed in terms of electron number density as $n_i=n_e/Z$. Further, $n_e$  can be expressed in terms of the Debye mass as $n_e=\mu m_D^2/3 e^2$, where, $m_D^2=e^2\mu^2/\pi^2$. 

Next we introduce a dimensionless variable  $y=q_z/q$,
\bea\label{eq:Ifi4}
I_{fi}&=&\frac{\mu m_D^2 eB}{3 Z e^2 (2\pi)^2 }\int dydq_x\frac{ dq_y }{16 \times 2}(n_f(p)-n_f(p'))|{\cal M}_{fi}|^2\nn\
\label{int_rate}
\eea
In order to calculate $|{\cal M}|^2$, we use following electronic spinor in presence of magnetic field, 
\begin{equation}\label{eq:up}
u(p) =\Bigg(\begin{matrix} 
\tilde\alpha \tilde AH_{n-1}(\xi) \\-s\tilde\alpha \tilde \beta \tilde H_n(\xi) \\s\tilde\beta \tilde A \tilde H_{n-1}(\xi) \\\tilde \beta \tilde B \tilde H_{n}(\xi)
 \end{matrix}
 \Bigg).
\end{equation}
Using above spinors and  the expression for photon propagator( Eq.(\ref{PiT})) in Eq.(\ref{mat_amp2})  one obtains,
\bea\label{intmatrix}
\sum_{spin} |{\cal M}|^2&=&(4\pi Ze^2)^2[\frac{2\pi ym_D^2}{(q_{\perp}^2+Re \Pi_L)^2+Im \Pi_L^2}-\frac{2\pi y m_D^2v_f^2}{2(q_{\perp}^2+Re \Pi_T)^2+Im \Pi_T^2)}]\nn\\
&&[ss'\tilde \alpha^2+\tilde \beta^2][ss'\tilde A \tilde A'I_{n'-1}I_{n-1}(q_{\perp})+\tilde A \tilde A'I_{n'-1}I_{n-1}(q_{\perp})]^2
\label{mat_amp2},
\eea 
where, $s$ and $s'$ are $\pm$, 
\begin{equation}\label{eq:alphabeta}
 \Bigg(\begin{matrix} 
\tilde \alpha \\ \tilde \beta\\
 \end{matrix}
 \Bigg)=\Bigg(\begin{matrix} 
\sqrt{\frac{1}{2}(1+\frac{mc^2}{\epsilon})}\\ \sqrt{\frac{1}{2}(1-\frac{mc^2}{\epsilon})}\\
 \end{matrix}
 \Bigg),
\end{equation}
\begin{equation}\label{eq:ABmatrix}
 \Bigg(\begin{matrix} 
\tilde A\\ \tilde  B\\
 \end{matrix}
 \Bigg)=\Bigg(\begin{matrix} 
 [\frac{1}{2}(1+\frac{sp_zc}{\sqrt{\epsilon^2-m^2c^4}})]^{1/2}\\
[\frac{1}{2}(1-\frac{-p_zc}{\sqrt{\epsilon^2-m^2c^4}})]^{1/2}\\ 
 \end{matrix}
 \Bigg)
\end{equation}
and
\bea\label{eq:hermitepol1}
I_{n',n}=\int_{-\infty}^{\infty} exp(iq_y y)\tilde H_{n'}(\xi')\tilde H_n(\xi)dy,
\eea
\bea\label{eq:hermitepol2}
H_n(\xi)=\frac{m\omega_B}{\pi \hbar}^{\frac{1}{4}}(2^n n!)^{\frac{-1}{2}}exp^{\frac{-\xi^2}{2}}H_n(\xi)
\eea
$H_n(\xi)$ is the Hermite polynomial, $\xi=\sqrt{m\omega_B}$.
Inserting above expressions in Eq.(\ref{eq:ABmatrix}) we obtain,

\bea
\sum_{spin} |{\cal M}|^2 &=&(4\pi Ze^2)^2[\frac{2\pi ym_D^2}{(q_{\perp}^2+Re \Pi_L)^2+Im \Pi_L^2}-\frac{2\pi y m_D^2v_f^2}{2(q_{\perp}^2+Re \Pi_T)^2+Im \Pi_T^2)}]\nn\\
&&[1+ss'
+\frac{m^2c^4}{\epsilon^2}(1-ss'){[1+ss'-
\frac{1}{2}\frac{ss'c^2}{\epsilon^2-m^2c^4}(\eta'p_{n'}-p_z)^2]
[F^2_{n',n}(q_{\perp})+F^2_{n'-1, n-1}(q_{\perp})]}]\nn\\
&&-\frac{ss'u\hbar \omega_B mc^2}{\epsilon^2-m^2c^4}
[F^2_{n'-1, n}(q_{\perp})+F^2_{n', n-1}(q_{\perp})]-\frac{sp_zc+s'\eta'p_n'c}{\sqrt{\epsilon^2-m^2c^4}}[F^2_{n', n}(q_{\perp})-F^2_{n'-1, n-1}(q_{\perp})]
\label{mat_amp3}
\eea
where, $\eta'=\pm$, $F_{n',n}=exp^{-u/2}u^{\frac{n-n'}{2}}\sqrt{\frac{n'!}{n!}}L_{n'}^{n-n'}$ and $u=\frac{1}{2m\omega_B}(q_x^2+q_y^2)$. The functions  $L_{n'}^{n-n'}(u)$ are Laguerre polynomials and  $F_{n',n}(u)$ are normalized as $\int_0^{\infty}F_{n',n}^2 du=1$.


Now, to perform the  integration in $y$ in Eq.(\ref{int_rate}) we use the following sum rule,
\bea\label{eq:sumrule}
  &&\int_{-1}^{1} \frac{dy}{y}\, \frac{1}{2\pi}\, 
  \frac{2\,{\rm Im}\,\Pi_{L}(y)}
       {(\bq^2{+}{\rm Re}\,\Pi_{L}(y))^2 + ({\rm Im}\,\Pi_{L}(y))^2}-\frac{2\,{\rm Im}\,\Pi_{T}(y)}
       {(\bq^2{+}{\rm Re}\,\Pi_{T}(y))^2 + ({\rm Im}\,\Pi_{T}(y))^2}\nn\\
&& =\left( \frac{1}{\bq^2+{\rm Re}\,\Pi_{T,L}(y{=}\infty)}
        -\frac{1}{\bq^2+{\rm Re}\,\Pi_{T,L}(y{=}0)}\right) 
        \eea
      
 In the limiting case, ${\rm Re}\,\Pi_{T,L}(y{=}\infty) = \frac{m_D^2}{3},
 {\rm Re}\,\Pi_{T}(y{=}0)=0, 
 {\rm Re}\,\Pi_{L}(y{=}0)=m_D^2$.

Using the above relations the interaction rate becomes,
\bea\label{Ififinal}
I_{fi}&=&\frac{n_i}{(2\pi)^2 }\int  dq_x\frac{ dq_y }{16\times 2}(n_f(p)-n_f(p'))\left[\frac{2}{3(q_{\perp}^2+\frac{m_D^2}{3})(q_{\perp}^2+m_D^2)}-\frac{v_f^2}{6q_{\perp}^2(q_{\perp}^2+\frac{\mu^2}{3})}\right]{\cal F},
\eea
where,
\bea\label{eq:F}
&&{\cal F}=\frac{4\pi \sigma_0}{m^2}[1+ss'
+\frac{m^2c^4}{\epsilon^2}(1-ss'){[1+ss'-
\frac{1}{2}\frac{ss'c^2}{\epsilon^2-m^2c^4}(\eta'p_{n'}-p_z)^2]
[F^2_{n',n}(u)+F^2_{n'-1, n-1}(u)]}]\nn\\
&&-\frac{ss'u\hbar \omega_B mc^2}{\epsilon^2-m^2c^4}
[F^2_{n'-1, n}(u)+F^2_{n', n-1}(u)]-\frac{sp_zc+s'\eta'p_n'c}{\sqrt{\epsilon^2-m^2c^4}}[F^2_{n', n}(u)-F^2_{n'-1, n-1}(u)].
\eea
We change the variable $q_y$ to $u$ and perform the integration as follows,
\bea\label{eq:qintegral}
\int dq_xdq_y&=& m\omega_B \int \frac{dq_x du}{\sqrt{2m\omega_Bu-q_x^2}}\nn\\
&=&m\omega_B \pi\int du
\eea
Finally, the particle scattering rate becomes,
\bea\label{eq:Ififinal2}
I_{fi}&=&\frac{n_i }{2}\int  du(n_f(p)-n_f(p'))[\frac{1 }{3(u+\frac{\xi}{3})(u+\xi)}-\frac{v_f^2}{6u(u+\frac{\xi}{3}}]{\cal F}.
\end{eqnarray}


\end{document}